\RequirePackage{fix-cm}

\documentclass[smallcondensed,final]{svjour3}     %
\smartqed  %

\usepackage{graphicx}
\usepackage{mathptmx}
\usepackage{bm}
\usepackage{subfigure}
\usepackage{caption} %
\usepackage{xcolor}
\usepackage{amsmath}               %
\usepackage{multicol,multirow,bigdelim} %
\usepackage[bottom]{footmisc}    %
\usepackage[utf8]{inputenc}          %
\usepackage[T1]{fontenc}
\usepackage{xparse}                     %
\usepackage[hidelinks]{hyperref}
\hypersetup{
    colorlinks,
    linkcolor={blue!80!black},
    citecolor={red!50!black},
    urlcolor={blue!50!black}
}
\usepackage{grffile} %
\usepackage{xstring} %
\usepackage{xspace} %
\usepackage{amssymb}
\usepackage{siunitx}

\usepackage{jabbrv}
\usepackage[numbers,sort&compress]{natbib}
\usepackage{pgfplots}
\usepackage{tikzscale} %
\usepgfplotslibrary{patchplots}
\usetikzlibrary{plotmarks}
\usetikzlibrary{arrows.meta}
\pgfplotsset{compat=newest} %
\pgfplotsset{every node/.style={inner sep=0,outer sep=0}}
\pgfplotsset{scale ticks below exponent={-5}}
\pgfplotsset{scale ticks above exponent={6}}
\newlength{\figureheight}
\newlength{\figurewidth}
\newlength{\defaultfigheight}
\newlength{\defaultfigwidth}

\journalname{Acta Mechanica}

\newcommand{\bcf}[1][]{
\setlength{\figurewidth}{\defaultfigwidth} 
\setlength{\figureheight}{\defaultfigheight} 
\ifx &#1&
\begin{figure}\centering 
\else
\begin{figure}[#1]\centering 
\fi}
\newcommand{\ecf}{\end{figure}}
\newcommand{\incg}[2][]{
\StrCut{#2}{.tikz}{\extensionlessfname}{\ignoredextension}
\ifx &#1& 
\def\svgwidth{\defaultfigwidth} 			
\includegraphics[width=\figurewidth]{#2} 	
\else 
\def\svgwidth{#1} 					
\includegraphics[width=#1]{#2} 			
\fi}

\newcommand{\beq}{\begin{eqnarray}}
\newcommand{\eeq}{\end{eqnarray}}
\newcommand{\bal}{\begin{align}}
\newcommand{\eq}[1]{\begin{eqnarray} #1 \end{eqnarray}}
\newcommand{\al}[1]{\begin{align} #1 \end{align}}

\renewcommand{\vec}[1]{{\bf{#1}}}

\newcommand{\rom}[1]{\textup{\uppercase\expandafter{\romannumeral#1}}} 
\newcommand{\tx}[1]{\text{#1}}

\newcommand{\refs}[1]{Sect.~\ref{sec:#1}\xspace}

\NewDocumentCommand{\reff}{mg}{Fig.~\ref{fig:#1}\IfNoValueF{#2}{ and Fig.~\ref{fig:#2}}\xspace}
\NewDocumentCommand{\refe}{mg}{Eq.~\ref{eq:#1}\IfNoValueF{#2}{ and Eq.~\ref{eq:#2}}\xspace}

\newcommand{\glosm}[3]{\newglossaryentry{#1}{name={\ensuremath{#2}},description={#3}}}

\usepackage[acronym,nonumberlist]{glossaries}
\makeglossaries
\setglossarystyle{super}
\setacronymstyle{long-short}

\glosm{u0}{u_0}{free stream velocity}
\glosm{d99}{\delta_{99}}{boundary layer thickness $u(y=\gls{d99})=0.99\,\gls{u0}$}
\glosm{d99n}{\delta_{99,neck}}{boundary layer thickness \gls{d99} at the neck center (\reff{setup})}
\glosm{Stbl}{St_{TBL}}{Strouhal number of the TBL $\omega\,\gls{d99}/\gls{u0}$}
\glosm{Sneck}{St_{neck}}{Strouhal number of the neck $\omega\,L_{x,neck}/\gls{up}$ (\refe{KHwavenumberSrBeta})}
\glosm{up}{u_+}{vortex sheet convection velocity (\refe{updef})}
\glosm{beta}{\beta}{frequency dependent tuning of the opening impedance amplitude (\refe{umdef})}
\glosm{tauw}{\tau_w}{wall shear stress $\rho_w \nu\, d/dy<u_x>|_{y=y_w}$}
\glosm{utau}{u_\tau}{wall friction velocity $\sqrt{\gls{tauw}/\rho_w}$}
\glosm{dnu}{\delta_\nu}{viscous lengthscale $\nu/\gls{utau}$}
\glosm{Retau}{Re_\tau}{friction Reynolds number $\gls{utau} \gls{d99} / \nu=\gls{d99} / \gls{dnu}$}
\glosm{Z}{Z}{complex-valued acoustic impedance $\gls{Z}=(\gls{r}+i k \gls{l})\gls{Y}$}
\glosm{r}{r}{acoustic resistance (excitation or damping)}
\glosm{l}{l}{acoustic reactance (length correction)}
\glosm{v}{v}{acoustic volume flux}
\glosm{Y}{Y}{characteristic impedance $Y=S/\rho c$ of a duct with the constant cross-section $S$}

\begin{document}
\title{An Acoustic Model of a Helmholtz Resonator under a Grazing Turbulent Boundary Layer}

\author{Lewin Stein \and Jörn Sesterhenn}

\institute{Lewin Stein \at
              Institut für Str\"omungsmechanik und Technische Akustik, M\"uller-Breslau-Str. 15 10623 Berlin \\
              \email{Lewin.Stein@tu-berlin.de}           %
              ORCID: 0000-0002-4298-2001 \\
           \and
           Jörn Sesterhenn \at
              Institut für Str\"omungsmechanik und Technische Akustik, M\"uller-Breslau-Str. 15 10623 Berlin
}

\date{Received: 11 July 2018 / Accepted: 11 November 2018}

\maketitle

\begin{abstract}

Acoustic models of resonant duct systems with turbulent flow depend on fitted constants based on expensive experimental test series.

We introduce a new model of a resonant cavity, flush-mounted in a duct or flat plate, under grazing turbulent flow.
Based on previous work by Goody, Howe and, Golliard, we present a more universal model where the constants are replaced by physically significant parameters.
This enables the user to understand and to trace back how a modification of design parameters (geometry, fluid condition) will affect acoustic properties.

The derivation of the model is supported by a detailed three dimensional Direct Numerical Simulation as well as an experimental test series.
We show that the model is valid for low Mach number flows ($M=0.01$-$0.14$) and for low frequencies (below higher transverse cavity modes).
Hence, within this range, no expensive simulation or experiment is needed any longer to predict the sound spectrum.
In principle the model is applicable to arbitrary geometries: Just the provided definitions need to be applied to update the significant parameters.
Utilizing the lumped element method, the model consists of exchangeable elements and guarantees a flexible use.
Even though the model is linear, resonance conditions between acoustic cavity modes and fluid dynamic unstable modes are correctly predicted.

\end{abstract}

\setcounter{tocdepth}{2}

\newpage
\tableofcontents

\printglossaries

\section{Introduction}

\subsection{Motivation}
\label{sec:motiv}

Typically, when gases (compressible fluids) stream along a hollow space, acoustic and turbulent flow strongly couple.
Many examples can be listed in which this coupling is of greatest importance.
Noise silencers consisting of cavity arrays are installed in most duct systems, in which tonal noise (due to a constant operating frequency) needs to be reduced, among others:
Air conditioning systems, ventilation plants, combustion engines.
Beside silencing properties, cavities may give rise to desirable tones of wind instruments like transverse flutes and organs or undesired `window buffeting' of moving vehicles. Acoustic cavity resonances may even cause severe material damage for instance in pipeline intersections.

\subsection{Research Goal: Acoustic Model}
\label{sec:intro}

Cavities under grazing turbulent flow are commonly surveyed \citep{Kook2002,golliardNoiseHelmholtzresonator02,Hemon2004}, mostly for industrial applications.
In practice, series of many different cavity configurations are experimentally tested for a certain design goal,
e.g. noise cancellation in \cite{dequandHelmholtzLikeResonator03}.
An understanding of the underlying physical processes, especially of the acoustic turbulence interaction is missing.
Such, there is a lack of easy applicable but realistic models, which are not dependent on expensive parameter studies.

\emph{In this work, we derive a new model of a Helmholtz resonator, which simplifies the design process and which is widely applicable due to a modular principle (Lumped-Element Method)}.
By Helmholtz resonator, we mean an almost closed cavity except for a neck opening (s. \refs{case}).
The Helmholtz resonator is excited by a grazing turbulent boundary layer (TBL) flow:
In the present case, the thickness of the TBL is smaller than the streamwise extension of the neck, $\gls{d99n}<L_{x-neck}$.
As a consequence, an unstable shear layer arises and thus a strong turbulence-acoustic interaction at the neck area is expected.
Particularly, the new model will be specialized in this turbulence-acoustic interaction (s. \refs{howe}), in contrast to common, purely acoustical descriptions.
\emph{The novelty of the model is characterized by its refined parameters.
The parameters are no longer meaningless fit parameters
but replaced by clearly defined quantities with a physical meaning.}
Thereby, we focus especially on the convection velocity, which is the key factor to model the acoustic impedance of a neck with a shear layer (\refs{uconv}).
The model is optimized for a low Mach number flow and low-frequency acoustics, which are typical operating conditions of duct systems.

In the following, previously established models are briefly surveyed.
Due to missing numerical studies (s. \refs{case}), they are predominantly based on experiments.
Often, the resonance frequencies but not the amplitudes are predicted:
An acoustic feedback mechanism of Kelvin-Helmholtz waves at the neck is described by \cite{Rossiter1964};
\cite{Tam2006} studied the additional coupling of these Kelvin-Helmholtz waves at the neck with the acoustic modes of an open cavity (valid for Mach numbers $M>0.2$). 
Models which can predict the sound pressure level (SPL), too, usually are either based on severe theoretical simplifications \citep{howeInfluenceMean81,howeTheoryUnsteady81}
or rely on empirical fits with a limited validity \citep{golliardNoiseHelmholtzresonator02}.
\\

This work is organized as follows:
\refs{case} introduces the geometry and conditions of our reference DNS.
In \refs{model} the new Helmholtz resonator model is derived in three steps:
\refs{basicModel} establishes the fundamental model structure.
\refs{source} implements the TBL as an acoustic source term into the model.
Based on theoretical works, \refs{Zflow} generally redefines a model of the turbulence-acoustic interaction.
The physical meaning and the validity of the model components is explained in \refs{results}.
A final model summary is made in \refs{model_sum}.
To conclude, \refs{end} discusses the impact of the new model.

\newpage
\section{Reference Case: DNS of a Helmholtz Resonator under Grazing Turbulent Flow}
\label{sec:case}

\bcf[b]
\def\svgwidth{0.6\textwidth} 
\begingroup%
  \makeatletter%
  \providecommand\color[2][]{%
    \errmessage{(Inkscape) Color is used for the text in Inkscape, but the package 'color.sty' is not loaded}%
    \renewcommand\color[2][]{}%
  }%
  \providecommand\transparent[1]{%
    \errmessage{(Inkscape) Transparency is used (non-zero) for the text in Inkscape, but the package 'transparent.sty' is not loaded}%
    \renewcommand\transparent[1]{}%
  }%
  \providecommand\rotatebox[2]{#2}%
  \ifx\svgwidth\undefined%
    \setlength{\unitlength}{669.07498169bp}%
    \ifx\svgscale\undefined%
      \relax%
    \else%
      \setlength{\unitlength}{\unitlength * \real{\svgscale}}%
    \fi%
  \else%
    \setlength{\unitlength}{\svgwidth}%
  \fi%
  \global\let\svgwidth\undefined%
  \global\let\svgscale\undefined%
  \makeatother%
  \begin{picture}(1,0.86044165)%
    \put(0.13510279,0.03598303){\makebox(0,0)[lb]{\smash{$x$}}}%
    \put(0.6616454,0.049){\makebox(0,0)[lb]{\smash{$10.8\delta$}}}%
    \put(0.73,0.23533347){\makebox(0,0)[lb]{\smash{$15.5\delta$}}}%
    \put(0.84211017,0.40548599){\makebox(0,0)[lb]{\smash{$12.6\delta$}}}%
    \put(0.42704116,-0.005){\makebox(0,0)[b]{\smash{$10.8\delta$}}}%
    \put(0,0){\includegraphics[width=\unitlength,page=1]{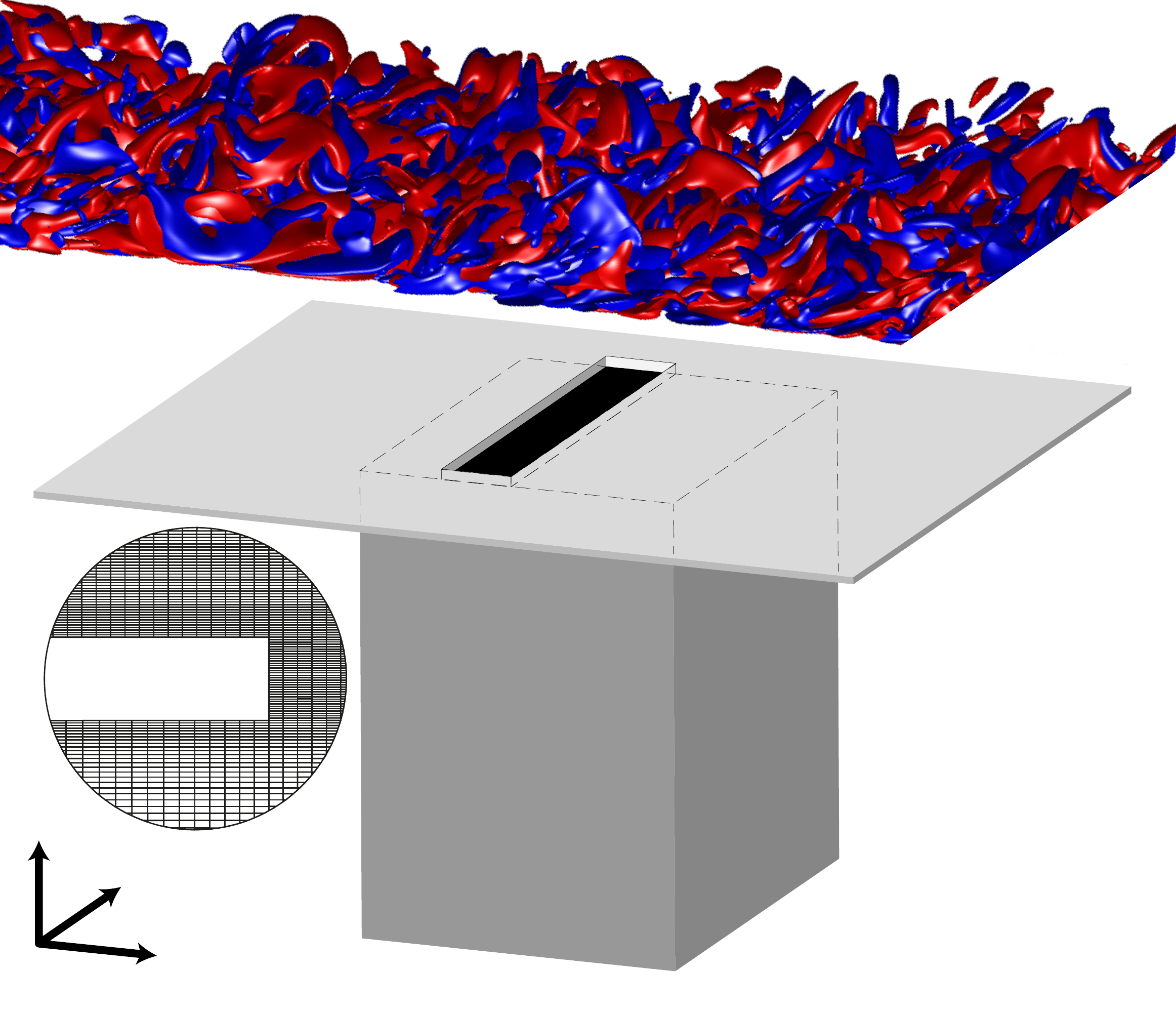}}%
    \put(0.02,0.16){\makebox(0,0)[lb]{\smash{$y$}}}%
    \put(0.11,0.11){\makebox(0,0)[lb]{\smash{$z$}}}%
    \put(0,0){\includegraphics[width=\unitlength,page=2]{Fig1input.pdf}}%
    \put(0.21821767,0.26896205){\makebox(0,0)[rb]{\smash{$0.1\delta$}}}%
    \put(0,0){\includegraphics[width=\unitlength,page=3]{Fig1input.pdf}}%
    \put(0.42,0.40621634){\makebox(0,0)[b]{\smash{$1.5\delta$}}}%
    \put(0,0){\includegraphics[width=\unitlength,page=4]{Fig1input.pdf}}%
  \end{picture}%
\endgroup
\caption{Reference case setup of a Helmholtz resonator under grazing turbulent flow.
At the top, vorticity isosurfaces ($\pm\SI{3000}{Hz}$ colored blue and red) of the TBL are visible.
Below the flat-plate flow, the rectangular cavity of the resonator is mounted.
All units are given in terms of the boundary layer thickness, defined at the neck center $\gls{d99n}=\SI{9.28}{mm}$.
Around the neck, the grid is further refined as indicated by the zoom window.}
\label{fig:setup}
\ecf

This section introduces the setup of the reference case, in particular how it was obtained by means of a Direct Numerical Simulation (DNS).
More details are published in \cite{steinNumericalSimulation18}.
A sketch of the geometry is provided in \reff{setup}.
In streamwise $x$-direction, a zero-adverse-pressure-gradient
turbulent boundary layer streams
over a rectangular cavity, which is flush-mounted inside the bottom wall.
The cavity is connected via a rectangular neck (aka opening) with the flat-plate.
The reference geometry is motivated by \citet{golliardNoiseHelmholtzresonator02}.
Experimentally \citeauthor{golliardNoiseHelmholtzresonator02} investigated a Mach number range from 0.01 to 0.14 and $\gls{d99n}/L_{x-neck}$ ratios (boundary layer thickness to streamwise neck length) from 0.7 to 5.5. This allows comparisons in \refs{range}.
We conducted one DNS at a Mach number of $M \cong 0.11$ and at $\gls{d99n}/L_{x-neck} \cong 0.7$.

To our knowledge, we conducted for the first time a DNS of this setup without simplification of the compressible Navier-Stokes equations.
In case a cavity with a neck is considered (Helmholtz resonator) the inflowing TBL is often missing \citep{tamComputationalExperimental05,tamComputationalExperimental10,rocheAircraftFan09}
or not all system scales are resolved, but some form of turbulence model is assumed \citep{eldredgeNumericalInvestigation07}. 
The geometrically closest match is the simulation of a cylindric cavity by \cite{Roche2010c}.
An acoustic solver (Lighthill's analogy) with source terms given by an incompressible fluid flow cannot be used due to the expected non-linear coupling of the TBL and the sound field in the neck region.

In order to simulate \SI{30}{ms} of physical time, we invested $7 \times 10^6$ CPU-hours with a grid composed of $1.2 \times 10^9$ grid points.
Due to the computational cost, a parameter study is only experimentally feasible.
But in contrast to an experiment, our single DNS run supplies a dataset of all flow variables at all times and spaces (20 terabytes were stored.).
We will evaluate this reference dataset to identify major and minor mechanisms and to legitimate our model assumptions.

\section{The New Helmholtz Resonator Model}
\label{sec:model}

In the following \refs{basicModel} the basic model structure of a Helmholtz resonator with a TBL flow is introduced.
The model consists of acoustical elements and flow-related elements.
The focus in this paper is to newly derive all flow-related elements
(purely acoustical elements stem from \citep{golliardNoiseHelmholtzresonator02}).
The two flow-related elements are
the TBL, which serves as a broadband source term for the new model (selected in \refs{source}),
and the most important flow-acoustic interaction element of the neck (defined in \refs{Zflow}).

\subsection{Basic Model Structure}
\label{sec:basicModel}

\begin{figure}[b]
\centering
\def\svgwidth{0.58\textwidth} 
\begingroup%
  \makeatletter%
  \providecommand\color[2][]{%
    \errmessage{(Inkscape) Color is used for the text in Inkscape, but the package 'color.sty' is not loaded}%
    \renewcommand\color[2][]{}%
  }%
  \providecommand\rotatebox[2]{#2}%
  \ifx\svgwidth\undefined%
    \setlength{\unitlength}{302.05130768bp}%
    \ifx\svgscale\undefined%
      \relax%
    \else%
      \setlength{\unitlength}{\unitlength * \real{\svgscale}}%
    \fi%
  \else%
    \setlength{\unitlength}{\svgwidth}%
  \fi%
  \global\let\svgwidth\undefined%
  \global\let\svgscale\undefined%
  \makeatother%
  \begin{picture}(1,0.86362271)%
    \put(0,0){\includegraphics[width=\unitlength,page=1]{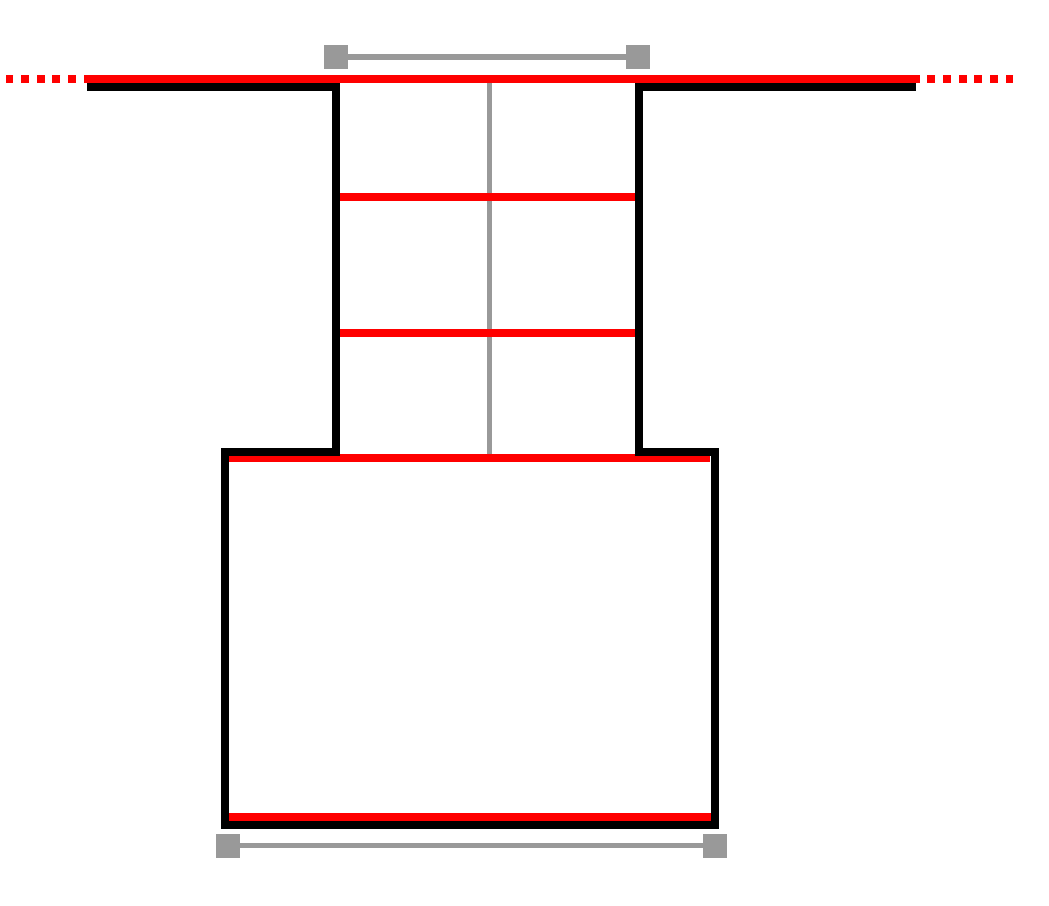}}%
    \put(0,0){\includegraphics[width=\unitlength,page=2]{Fig2input.pdf}}%
    \put(0,0){\includegraphics[width=\unitlength,page=3]{Fig2input.pdf}}%
    \put(0,0){\includegraphics[width=\unitlength,page=4]{Fig2input.pdf}}%
    \put(0,0){\includegraphics[width=\unitlength,page=5]{Fig2input.pdf}}%
    \put(0,0){\includegraphics[width=\unitlength,page=6]{Fig2input.pdf}}%
    \put(0,0){\includegraphics[width=\unitlength,page=7]{Fig2input.pdf}}%
    \put(0,0){\includegraphics[width=\unitlength,page=8]{Fig2input.pdf}}%
    \put(0.04234521,0.82556769){\makebox(0,0)[lb]{\smash{\textbf{flat plate}}}}%
    \put(0.0396973,0.71582553){\makebox(0,0)[lb]{\smash{\textbf{infinite flange}}}}%
    \put(0.04134941,0.59743809){\makebox(0,0)[lb]{\smash{\textbf{flow interaction}}}}
    \put(0.04110047,0.4790506){\makebox(0,0)[lb]{\smash{\textbf{area jump}}}}%
    \put(0.04096467,0.26199984){\makebox(0,0)[lb]{\smash{\textbf{cavity}}}}%
    \put(0.65193282,0.59379886){\color[rgb]{0,0,0}\makebox(0,0)[lb]{\smash{\textbf{$
    	M_{neck}
    	$}}}}%
    \put(0.7,0.23894293){\color[rgb]{0,0,0}\makebox(0,0)[lb]{\smash{\textbf{$
	M_{cavity}
	$}}}}%
    \put(0.39366278,0.71150288){\color[rgb]{0,0,0}\makebox(0,0)[b]{\smash{\textbf{$l_{rad}$}}}}%
    \put(0.39347667,0.59254964){\color[rgb]{0,0,0}\makebox(0,0)[b]{\smash{\textbf{$l_{flow}$}}}}%
    \put(0.39380183,0.4790506){\color[rgb]{0,0,0}\makebox(0,0)[b]{\smash{\textbf{$l_{jump}$}}}}%
    \put(0.53964882,0.71150288){\color[rgb]{0,0,0}\makebox(0,0)[b]{\smash{\textbf{$r_{rad}$}}}}%
    \put(0.53946272,0.59254964){\color[rgb]{0,0,0}\makebox(0,0)[b]{\smash{\textbf{$r_{flow}$}}}}%
    \put(0.53978787,0.4790506){\color[rgb]{0,0,0}\makebox(0,0)[b]{\smash{\textbf{$$}}}}%
    \put(0.46777713,0.82738934){\color[rgb]{0,0,0}\makebox(0,0)[b]{\smash{\textbf{$S_{neck}$}}}}%
    \put(0.46374957,0.01472036){\color[rgb]{0,0,0}\makebox(0,0)[b]{\smash{\textbf{$S_{cavity}$}}}}%
    \put(0.73,0.81){\color[rgb]{0,0,0}\makebox(0,0)[lb]{\smash{\textbf{$y_{top}=0$}}}}%
    \put(0.70593369,0.08){\color[rgb]{0,0,0}\makebox(0,0)[lb]{\smash{\textbf{$y_{bot}=-L_{y-neck}-L_{y-cavity}$}}}}%
    \put(0.4655228,0.11){\color[rgb]{0,0,0}\makebox(0,0)[b]{\smash{\textbf{$v_{bot}=0$}}}}%
    \put(0.03958737,0.13741535){\color[rgb]{0,0,0}\makebox(0,0)[lb]{\smash{\textbf{$y$}}}}%
    \put(0.11468616,0.05857205){\color[rgb]{0,0,0}\makebox(0,0)[b]{\smash{\textbf{$x$}}}}%
  \end{picture}%
\endgroup
\caption{Basic model structure of the Helmholtz resonator model using the Lumped-Element Method.
To provide a clear view of the neck elements the figure proportions are distorted (for corrects proportions s. \reff{setup}).
The red horizontal lines mark the start and end surfaces of the elements.
The cavity has a height $L_{y-cavity}$ and a constant cross-section $S_{cavity}$.
The cavity and the neck have the same spanwise depth $L_{z-cavity}=L_{z-neck}$.
}
\label{fig:lumpedsketch}
\end{figure}

Based on the Lumped-Element method \citep{munjalAcousticsDucts87}, the underlying basic model structure is
\beq
\begin{pmatrix}
p_{bot} \\
0
\end{pmatrix}
=&
\underbrace{
\begin{pmatrix}
	\cos( k L_{y-cavity} ) & i \sin( k L_{y-cavity} ) Y_{cavity} \\
	-i \sin( k L_{y-cavity} )/Y_{cavity} & \cos( k L_{y-cavity} )
\end{pmatrix}
}_{M_{cavity}}
\underbrace{
\begin{pmatrix}
1&Z_{neck} \\
0&1
\end{pmatrix}
}_{M_{neck}}
\begin{pmatrix}
p_{top}\\
v_{top}
\end{pmatrix}.
\label{eq:rawmod}
\eeq
As illustrated in \reff{lumpedsketch}, \refe{rawmod} relates the acoustic pressure $p_{top}$ and the acoustic volume flux $\gls{v}_{top}$ above the cavity (``top'' position)
with the acoustic pressure at the bottom of the cavity (``bot'' position with a hard wall i.e. $v_{bot}=0$).
The Lumped-Element method assumes linear one-dimensional plane harmonic waves with the relation $\omega=k c$ of frequency, wavevector, and speed of sound, respectively.
The standard transfer matrix $M_{cavity}$ describes acoustic wave propagation in $y$-direction inside the cavity with the constant cross-section $S_{cavity}$, the characteristic impedance $Y_{cavity}=S_{cavity}/\rho c$ and the cavity height $L_{y-cavity}$.
Emphasis is to be laid on the neck impedance $Z_{neck}$, which combines all effects related to the neck geometry and the flow.
Its detailed discussion follows.

Typical challenges in dealing with acoustically-resonant cavities are either to silence existing tonal noise or to prevent cavity resonances before they happen (\refs{motiv}).
In the first case the transfer or damping function of external acoustic waves passing by the cavity is seeked (i.e. $Z_{top}=p_{top}/v_{top}$ of \refe{rawmod}).
In the second case the penetration of acoustic waves into the cavity resonator is of interest (i.e. $p_{bot}/p_{top}$ of \refe{rawmod}).
As a proof of concept, this work discusses the second case exclusively,
the prediction of the sound pressure level (SPL) spectrum at the bottom of the cavity:
\beq
\Phi_{bot}(\omega)=&T(\omega)
\, \Phi_{top}(\omega)
\,,
\label{eq:GHGmodel}
\eeq
being $\Phi$ the power spectral density of pressure and $T=|p_{bot}/p_{top}|^2$ the transmission function.
Without constraints, the model elements,
explicitly derived in this paper,
can be directly applied in the first case, too.
The source term $\Phi_{top}$ of \refe{GHGmodel} will be specified in \refs{source}.

In the following, the transmission function $T$ is derived from \refe{rawmod}.
Flow related and purely acoustical effects are separated from each other by different impedance elements:
$Z_{neck}=Z_{jump}+Z_{flow}+Z_{rad}$.
\reff{lumpedsketch} qualitatively illustrates the horizontal positions of these three neck elements separated by red lines.
The separation brings the advantage of differentiating and investigating physical effects independently.
An individual element can be adapted to the specific application.
$Z_{jump}$ and $Z_{rad}$ are purely acoustical elements:
$Z_{jump}$ accounts for the cross-section jump between the cavity and the neck;
$Z_{rad}$ describes the radiation losses of a neck opening mounted in an infinitely extended plate (infinite flange).
All interactions of acoustic waves with the
shearing flow around the cavity opening
are incorporated in $Z_{flow}$.
We split the real and imaginary-valued part of each impedance element according to
$\gls{Z}=(\gls{r}+i k \gls{l})\gls{Y}$.
Altogether (analogous to \citep{golliardNoiseHelmholtzresonator02}) the transmission function for \refe{GHGmodel} derived from \refe{rawmod} is
\al{
T^{-1}                   &= \sin^2(k L_{y-cavity}) \left( \Lambda_l^2 + \Delta_r^2 \right),
\label{eq:HRtransfer} \tx{ being}\\
\Lambda_l &= \cot(k L_{y-cavity}) - S_{ratio} k (l_{jump} + l_{flow} + l_{rad})
\tx{ the effective length,}
\label{eq:dtot}
\\
\Delta_r                &= S_{ratio} (r_{flow}+r_{rad})
\tx{ the energy transfer,}
\label{eq:rtot}
}
and $S_{ratio}=S_{cavity}/S_{neck}$ the cross-section surface ratio.
$\Lambda_{l}$ can be related to the total effective length of the cavity including all length corrections.
Its zeros are the longitudinal angular $y$-wavenumbers $k=\omega/c$ of the cavity:
The first zero corresponds to the Helmholtz base frequency;
the second zero is the largest $y$-wavelength fitting into the cavity and so on.
The total resistance $\Delta_r$ can be interpreted as amplitude modulation of acoustic waves (sign dependent excitation or damping).
To determine the transmission function $T$ of \refe{HRtransfer}, its impedance elements $Z_{jump}$, $Z_{flow}$, $Z_{rad} $ must be specified first.
The purely acoustical elements (no-flow)
\al{
l_{jump} = L_{x-neck} \ln{\frac{2 L_{x-cavity}}{\pi L_{x-neck}}},
\quad 
r_{rad} = \frac{S_{neck} k^2}{2\pi},
\quad 
l_{rad} = \frac{L_{x-neck}}{\pi} \ln\left(\frac{8 L_{z-neck}}{e L_{x-neck}}\right)
\label{eq:howeUzeroD}
}
are adapted from \citet{golliardNoiseHelmholtzresonator02}.
The key point of this paper is to derive the flow related impedance $Z_{flow}=(r_{flow}+i k l_{flow})Y_{neck}$ fail-safe (s. \refs{Zflow}),
which is missing in the transmission function $T$ (\refe{dtot}{rtot}).

\subsection{Model Source Term of a Turbulent Boundary Layer}
\label{sec:source}

The examined Helmholtz resonator is driven by a TBL.
In this section, we select the source term $\Phi_{top}$ of the new model, that most realistically models the natural broadband excitation generated by a TBL.
First, we evaluate the SPL frequency spectrum of the DNS data.
Second, we adopt a validated generally usable model to the present case.

To set up the source term $\Phi_{top}(\omega)$ of \refe{GHGmodel} we need to determine the power spectrum of pressure fluctuations below an undisturbed TBL ($y=0$).
Because the TBL of the DNS described in \refs{case} is disturbed by the presence of the cavity,
in this section, we will evaluate a second DNS
without a cavity but with the same conditions otherwise.
Details of this undisturbed TBL simulation are specified in \cite{steinNumericalSimulation18}.
As an intermediate step, the $\Phi_{top}(k_x,\omega)$ power spectra of the DNS are calculated by a Discrete Fourier Transformation (DFT) in space and time (Wiener-Khinchin theorem, for details s. \cite{steinNumericalSimulation18}).
This more comprehensive $k_x-\omega$ representation is needed later in \refs{Zflow}.
As the spatial input of the DFT, two different top-hat windows ($L_{x-DFT}=1.5\,\gls{d99n},7.5\,\gls{d99n}$) in the streamwise direction of the TBL are selected (at the bottom wall $y=0$). %
The DFT output is averaged over all spanwise $z$ locations.
In this section (in contrast to \refs{Zflow}) the $k_x-\omega-$spectra are averaged over all $k_x$, where $k_x$ is the streamwise wavevector.
The two resulting narrowband SPL$(\omega)$ are displayed in \reff{goody}.
The evaluable DNS time window of \SI{20}{ms} results in a narrowband frequency bin of $f_\text{bin} = 1/T = \SI{50}{\Hz}$.

\bcf[b]
\setlength\figurewidth{0.8\textwidth}
\setlength\figureheight{0.45\textwidth}
\incg{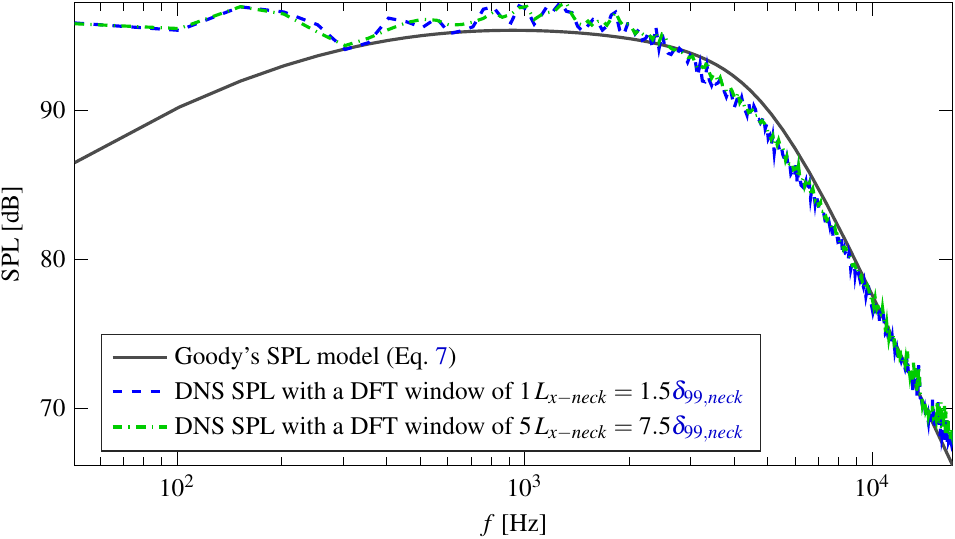}
\caption{Pressure fluctuations of a undisturbed TBL at the wall
($k_x$-averaged, $y=0$, $z$-averaged). Goody's Model is contrasted with the DNS results for different streamwise DFT windows.}
\label{fig:goody}
\ecf

To set up a generally applicable model source term, which is valid for different kinds of TBL flows,
we compared different
spectral models to our DNS spectra.
In conclusion, we found that the model of \citet{goodyEmpiricalSpectral04} is the best fit for the present TBL.
This is consistent with the review paper of \cite{hwangComparisonSemiempirical09}.
\citeauthor{goodyEmpiricalSpectral04}'s model features the typical Strouhal number scaling $\gls{Stbl}=\omega \gls{d99}/\gls{u0}$,
a prefactor $\rho^2_w \gls{tauw}^4 \gls{d99}/\gls{u0}$ like the %
model by \citet{chaseCharacterTurbulent87},
an correct exponential growth with $\omega^2$ at low frequencies,
an exponential decay with $\omega^{-5}$ at high frequencies,
and a wall friction Reynolds number \gls{Retau} dependency:
\beq
\Phi_{TBL}(\omega) =
\rho^2_w \gls{tauw}^4 \frac{ \gls{d99}}{\gls{u0}}
\frac{C\ \gls{Stbl}^2}{\left\{ \gls{Stbl}^{3/4}+0.5 \right\}^{3.7} + \left\{ 1.1 \, \gls{Stbl} \left( \gls{Retau} \gls{utau}/\gls{u0} \right)^{-0.57} \right\}^7}
\;.
\label{eq:goody}
\eeq

The lower index ``$w$'' denotes quantities defined at the wall.
In \reff{goody}, beside the DNS SPL, \refe{goody} is evaluated and displayed, too. %
Goody's model agrees with the DNS within an accuracy of $\pm\SI{2}{dB}$ for all frequencies above \SI{300}{Hz}.
Only the offset constant $C$ of \refe{goody} was increased from $3$ to $25$ ($+\SI{9}{dB}$) to fit the DNS data.
This is comparable to the increase needed to match the Chase model in \citep{golliardNoiseHelmholtzresonator02}.
Hence, the value of $C$ is likely to be valid in other similar cases, too.
The first four frequency bins are overestimated by the DFT, due to the noise levels of the short DNS time series available (\SI{20}{ms}, s. \refs{case}).
Ideally, to calculate a universal TBL spectrum, the streamwise input signal length $L_{x-DFT}$ DFT is as long as possible.
However, by comparing the two different $L_{x-DFT}$ lengths in \reff{goody}, no substantial differences occur.
The wavenumber filtering caused by the spatial confinement (i.e. finite neck opening $L_{x-neck}$) is negligible.
Therefore, the TBL spectrum by Goody's model can be utilized directly as a source term
of the Helmholtz resonator model
$\Phi_{top}=\Phi_{TBL}$,
which acts on the localized surface of the neck only.
Goody's model is generally usable without DNS because it is only dependent on universal TBL parameters
($\gls{d99}, \gls{u0}, \gls{Retau},\dots$).

\subsection{Flow-Acoustic Interaction of an Opening with a Vortex Sheet Redefined} %
\label{sec:Zflow}

This section elaborates on the central impedance element $Z_{flow}$ of the new Helmholtz resonator model: 
the incorporation of the interaction between the shearing flow around the cavity opening and the acoustic waves resonating inside the cavity.
So far the source term $\Phi_{top}$ of \refe{GHGmodel} describes the outer undisturbed TBL source, only.

In the next \refs{howe}, $Z_{flow}=(r_{flow}+i k l_{flow})Y_{neck}$ of the transmission function $T$ (\refe{HRtransfer}) is linked in a new way to \citeauthor{howeAcousticsFluidStructure98}'s theory \citep{howeAcousticsFluidStructure98}.
Finally, \refs{uconv} completely defines $Z_{flow}$.
The novelty of the new definition lies in a more general description.
Later, in \refs{reinterpretation}, a physical reinterpretation of the new definition is given.

\subsubsection{Conversion of Howe’s Rayleigh Conductivity into an Acoustic Impedance Element}
\label{sec:howe}

To legitimate the use of Howe's extensive theoretical work on the Rayleigh conductivity of apertures in turbulent flow
we evaluate the DNS data around the neck opening first.
The incoming boundary layer thickness is smaller than the streamwise neck length.
As a result, an unstable shear layer with Kelvin-Helmholtz waves arises inside the neck zone.
In \reff{domain_opening_vsheet}(b) vorticity isosurfaces with $\nabla \times \vec u = \pm\SI{3000}{Hz}$ are depicted. %
A secondary vortex sheet covering all the opening surface at constant height $y = 0$ can be visually distinguished from the dominant streaks of the incoming TBL.
For a better understanding, this vortex sheet is colored red and blue.
This secondary vortex sheet leads to a reattachment of the TBL on the wall for a small stretch downstream of the neck:
In \reff{domain_opening_vsheet}(b) for about one $L_{x-neck}$ downstream of the neck length no (white) spacing between the lowest TBL whirls and the plate is visible.
Inside the cavity, whirls are only present in the immediate vicinity of the opening.
The size ratio of the small neck and the large cavity is apparent in \reff{domain_opening_vsheet}(a).
The same is true for a snapshot at another time or for a different vorticity frequency than \SI{3000}{Hz}.
Inside the cavity, except the neck region, we observe nearly zero flow and acoustical phenomena dominate.

\bcf[t]
\begin{subfigure}[]{}
\def\svgwidth{3cm} 
\begingroup%
  \makeatletter%
  \providecommand\color[2][]{%
    \errmessage{(Inkscape) Color is used for the text in Inkscape, but the package 'color.sty' is not loaded}%
    \renewcommand\color[2][]{}%
  }%
  \providecommand\transparent[1]{%
    \errmessage{(Inkscape) Transparency is used (non-zero) for the text in Inkscape, but the package 'transparent.sty' is not loaded}%
    \renewcommand\transparent[1]{}%
  }%
  \providecommand\rotatebox[2]{#2}%
  \ifx\svgwidth\undefined%
    \setlength{\unitlength}{405.75bp}%
    \ifx\svgscale\undefined%
      \relax%
    \else%
      \setlength{\unitlength}{\unitlength * \real{\svgscale}}%
    \fi%
  \else%
    \setlength{\unitlength}{\svgwidth}%
  \fi%
  \global\let\svgwidth\undefined%
  \global\let\svgscale\undefined%
  \makeatother%
  \begin{picture}(1,1.24158967)%
    \put(0,0){\includegraphics[width=\unitlength,page=1]{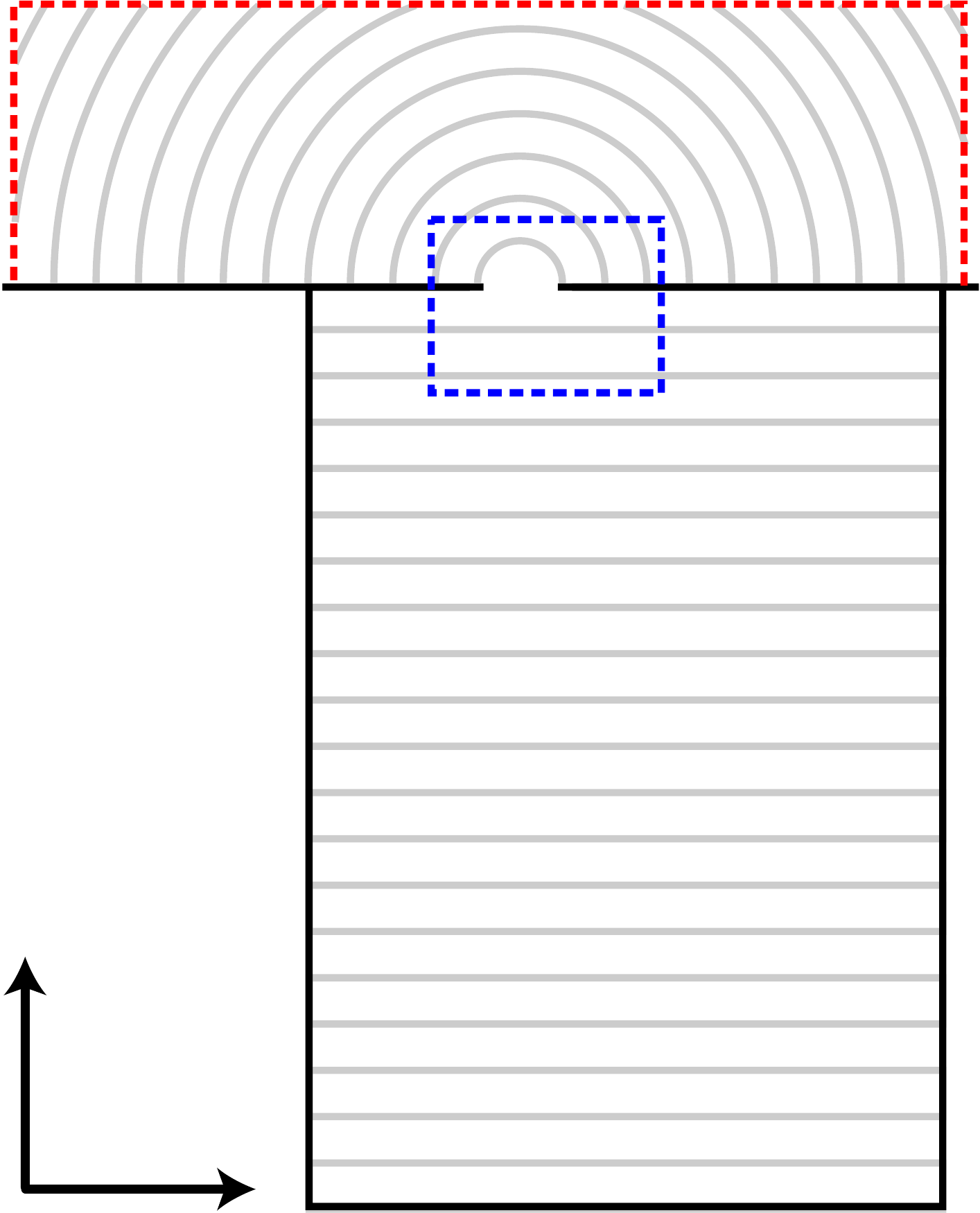}}%
    \put(0.06138113,0.20639425){\color[rgb]{0,0,0}\makebox(0,0)[lt]{\begin{minipage}{0.31992598\unitlength}\centering \textbf{x}\end{minipage}}}%
    \put(0.03162059,-0.13020243){\color[rgb]{0,0,0}\makebox(0,0)[lt]{\begin{minipage}{0.22320417\unitlength}\centering \end{minipage}}}%
    \put(-0.22692423,0.44454829){\color[rgb]{0,0,0}\makebox(0,0)[lt]{\begin{minipage}{0.55801053\unitlength}\centering \textbf{y}\end{minipage}}}%
    \put(0.03682666,0.58979762){\color[rgb]{0,0,0}\makebox(0,0)[lt]{\begin{minipage}{1.19424094\unitlength}\centering \textbf{cavity}\end{minipage}}}%
    \put(-0.25252665,1.2205458){\color[rgb]{0,0,0}\makebox(0,0)[lt]{\begin{minipage}{1.38889706\unitlength}\centering \textbf{plate}\end{minipage}}}%
  \end{picture}%
\endgroup
\end{subfigure}
\begin{subfigure}[]{}
\incg[8.3cm]{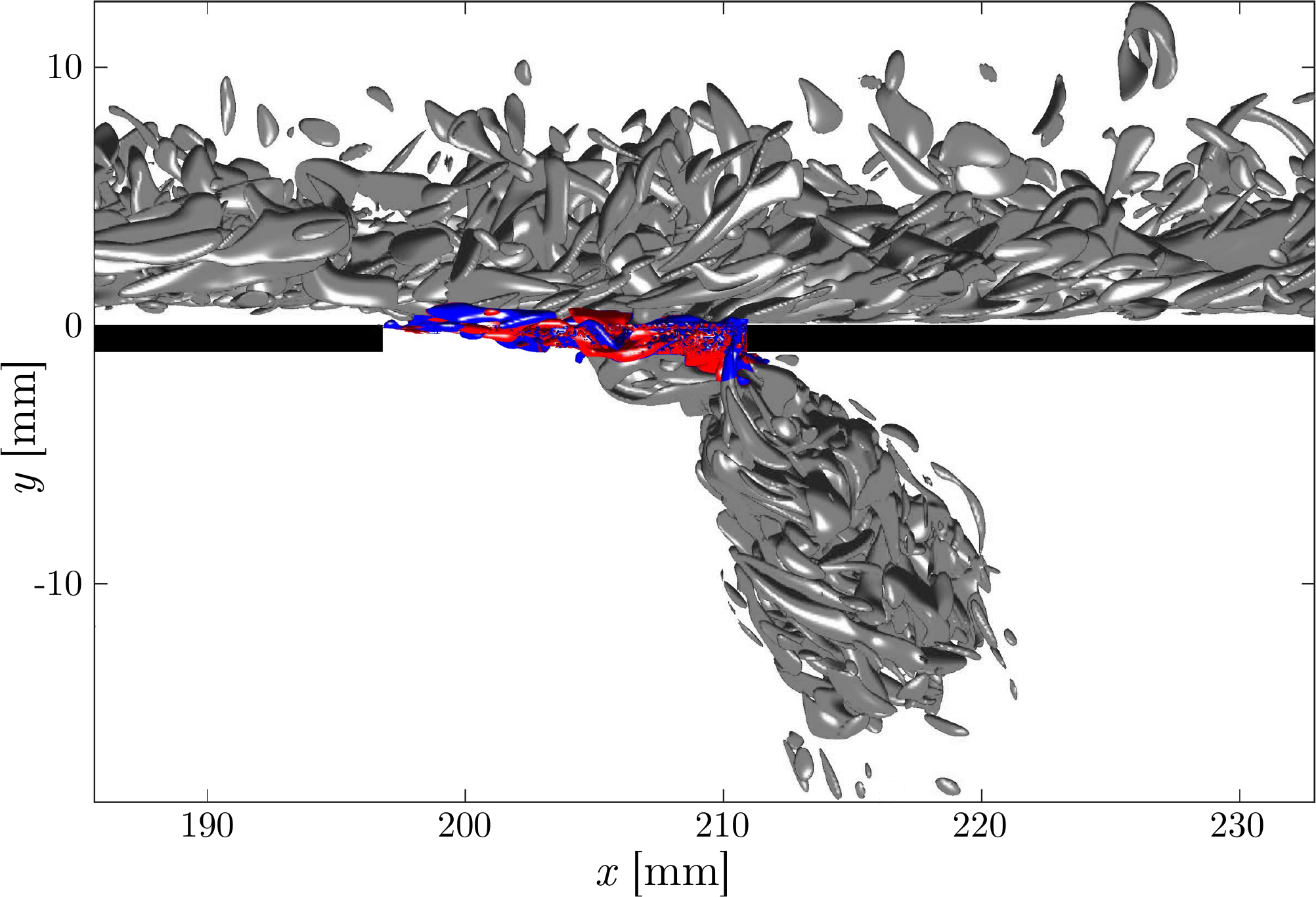}
\caption{DNS Simulation of the reference case (\refs{case}):
(a) calculation domain in exact proportions with hard walls (black lines) and nonreflecting boundary conditions (red dashed line),
(b) %
snapshot of two vorticity isosurfaces with $\pm\SI{3000}{Hz}$ (blue and red).
In order to highlight the vortex sheet at $y\approx0$, everything else is grayed out.
(b) shows a section near the neck only.
Its trim area is indicated by the dashed blue box of (a).
}
\label{fig:domain_opening_vsheet}
\end{subfigure}
\ecf

Based on these observations, the following assumptions for the Helmholtz resonator model are reasonable: %
Acoustic waves can propagate inside the cavity without any fluid related effects.
The crucial interaction of the acoustic waves with the turbulent flow is restricted to the vortex sheet of the opening.
The vortex sheet is very thin compared to the acoustic wavelengths of interest (cavity size and larger).

Under these assumptions, \citeauthor{howeAcousticsFluidStructure98}'s model of acoustic waves tunneling perpendicular through an opening with an infinitely thin vortex sheet (caused by grazing flow) can be used.
Howe models the local $y$-displacement $\zeta(x,z)$ of the vortex sheet dependent on
the frequency $\omega$ of the tunneling acoustic wave,
the $xz$-shape of the neck (not dependent on $L_{y-neck}$),
and two characteristic flow velocities $u_\pm$ above and below the thin vortex sheet.
\citet[Sec. 5.3.1]{howeAcousticsFluidStructure98} demonstrates, that his Rayleigh conductivity definition depends on the integrated vortex sheet displacement
\beq
\frac{K_R}{i \omega \rho}
\equiv
\frac{v}{p_+-p_-}=
\frac{\pi L_{x-neck}}{2} \int_{S_{neck}} \zeta(x,z) dxdz
\,,
\label{eq:KRdef}
\eeq
being $p_\pm$ the acoustic pressure above and below the opening,
and $v=v_+=v_-$ the total acoustic volume flux through the opening.
$K_R$ includes no-flow effects (due to the narrowing of the neck), too. %
In contrast to the present DNS case, Howe considers an opening inside an infinite extended and infinite thin wall.
Considering the left part of \refe{KRdef} as equivalent to the transfer matrix $M_{neck}(Z_{neck}=(r_{neck}+i k l_{neck})Y_{neck})$ of \refe{rawmod},
we derive the relation
\beq
\widetilde{l}_{neck}= -\Re \left[\frac{S_{neck}}{K_R}\right],
\tx{ and }
\widetilde{r}_{neck} = k\ \Im \left[\frac{S_{neck}}{K_R}\right].
\label{eq:ZflowKRrel}
\eeq
Positive reactance $\widetilde{r}_\text{neck}>0$ implies acoustic damping, which is equivalent to $\Im{[K_R]}<0$.
The tilde of $\widetilde{l}_{neck}$ and $\widetilde{r}_{neck}$ is needed, to denote Howe's different geometry.
Since only the impedance $Z_{flow}$ of the vortex sheet tunneling process itself %
is required to complete the new Helmholtz resonator model,
we simply subtract the dispensable no-flow part from both sides of \refe{ZflowKRrel}.
In that way, effects of Howe's infinite extended wall are excluded.
$Z_{flow}$ becomes
\beq
l_{flow}
=& \Re \left[\frac{S_{neck}}{K_{R_0}-K_{R}}\right]
\tx{ and }
r_{flow}
=& k\ \Im \left[\frac{S_{neck}}{K_{R}-K_{R_0}}\right]
\;,\label{eq:Zflow} %
\eeq
being $K_{R_0}= \lim_{\gls{u0} \to 0} K_{R}$ the Rayleigh conductivity in the case of an opening without flow nor vortex sheet.

To determine the Rayleigh conductivity $K_{R}$, the vortex sheet displacement $\zeta(x,z)$ of \refe{KRdef} needs to be computed.
For arbitrary $xz$-shapes of the opening, the vortex sheet displacement $\zeta(x,z)$ can be determined by numerically solving the conditional equation
\beq
 \int_S \frac{ \zeta(\vec x, \vec x')}{\vec x - \vec x'}  d\vec x' + \lambda_1(z)e^{i\sigma_1x} + \lambda_2(z)e^{i\sigma_2x} =1
 \label{eq:vortex_disp}
\eeq
for all locations $\vec x=(x,z)^T \in S_{neck}$ of the opening surface \citep{Grace1998a}.
$\sigma_{1,2}$ are Kelvin-Helmholtz wavenumbers dependent on the Strouhal number $\gls{Sneck}$ of the vortex sheet:
\eq{
\sigma_{1,2}= \frac{\gls{Sneck}}{2}\frac{1\pm i}{1\pm i \gls{beta}}
\,,\quad
\gls{Sneck}=\frac{\omega L_{x-neck}}{\gls{up}}
\,,\quad
\gls{beta}=\frac{u_-}{\gls{up}}
\label{eq:KHwavenumberSrBeta}
\,.
}
$\gls{beta}=1$ implies a real-valued wavenumber $\sigma_{1,2}$ scaling with $\gls{Sneck}$.
Only in case $\gls{beta}\neq1$ the wavenumber becomes complex, i.e. the Kelvin-Helmholtz waves $e^{i\sigma x}$ are amplified.
In case of simple geometries like a rectangular neck with $L_{z-neck} \gg L_{x-neck}$ as in the present case (cf. \reff{setup}), $\zeta(x,z)$ can be analytically determined
\citep{howeInfluenceMean81}.
The solution
\bal
K_R=&
\frac{\pi}{2} L_{z-neck} / \left\{ \mathcal{F}(\sigma_1,\sigma_2) + \Psi \right\}
\label{eq:howe}
\,,\\
\mathcal{F}(\sigma_1,\sigma_2)=&
\left\{ 
J_0(\sigma_1)f_1(\sigma_2)-J_0(\sigma_2)f_1(\sigma_1)
\right\} / \left\{
f_0(\sigma_2)f_1(\sigma_1)-f_0(\sigma_1)f_1(\sigma_2)
\right\}
\,,\notag\\
\Psi =& \ln(2) - \frac{\pi}{2} \left( a_- + a_+ \right)
\,\notag
\end{align}
will be utilized to evaluate $Z_{flow}$ of the Helmholtz resonator model for the present setup. %
$f_1(\sigma)=\sigma\{J_0(\sigma)-2 f_0(\sigma)\}$, $f_0(\sigma)=i\sigma\{J_0(\sigma)-iJ_1(\sigma)\}$ are shorthands of combined Bessel functions $J_{0,1}$.
The spatial dimensions are expressed as $a_\pm=\ln(e L_{x-neck}/4 L_{z-neck})/\pi$.
Without flow, the function $\lim_{\gls{up} \to 0} \mathcal{F} =\lim_{\gls{Sneck} \to \infty} \mathcal{F}=-2$ becomes real-valued and
there is no energy exchange of the acoustics with the vortex sheet: $\Im[S/K_{R_0}]=0$.

By now we derived all lumped impedance terms ($r$'s and $l$'s) of the transmission function $T$ (\refe{HRtransfer}) for the new Helmholtz resonator model (\refe{GHGmodel}).
With \refe{Zflow} we formally linked the impedance element $Z_{flow}=(r_{flow}+i k l_{flow})Y_{neck}$ to Howe's $K_R$.
However, both the analytical (\refe{howe}) and the numerical solution (\refe{vortex_disp}) of $K_R$ are dependent on two unknown constants:
the characteristic flow velocities $u_\pm$ (s. $\sigma_{1,2}$ of \refe{vortex_disp}).
Most users fit $u_\pm$ to match their particular case only. 
To solve this outstanding issue of Howe's theory,
we propose a general usable $u_\pm$ definition in the next \refs{uconv}.

\subsubsection{Convection Velocities in Howe's Theory: A Refined, Unique Definition}
\label{sec:uconv}

Below, we derive a novel, unique definition of the parameters $u_\pm$ of \refe{KHwavenumberSrBeta},
which are governing Howe’s Rayleigh conductivity of an opening, i.e. the impedance $r_{flow}$ and $l_{flow}$ (\refe{Zflow}).
Though the theory of Howe is widely-used, $u_\pm$ are unclearly defined so far.
Sometimes $u_\pm$ are referred to as a mean velocity \citep{howeAcousticsFluidStructure98}, sometimes interpreted as turbulent convection velocity \citep{peatAcousticImpedance03}.
This section begins with a definition of the general convection velocity $u_c$ (\refe{uconv}).
$u_c$ varies depending on the spatial position and frequency.
Then, the four-dimensional field $u_c$ is spectrally and spatially averaged to deduce the single parameter \gls{up} (\refe{updef}),
which best represents the neck vortex sheet velocity, within Howe's theory.
\gls{beta} is set as a constant (\refe{umdef}).
In doing so, \gls{up} and \gls{beta}
become universally applicable quantities %
and gain a physical meaning.\\

\textbf{General Convection Velocity} $\bm{u_c}$\quad The turbulent convection velocity is an ill-defined property, which describes the speed of vortices in general.
Depending on the application, many definitions exist \citep{blakeTurbulentBoundarylayer70,alamoEstimationTurbulent09},
both in real and spectral space.
The definitions range from broadband or group velocities to narrowband or phase velocities.
Sometimes the definitions depend on spatial (wavelength) and temporal (frequency) length-scales, sometimes averages are taken into account.

A common, robust definition of the narrow band convection velocity $u_c$ is the convective ridge maximum of the SPL $k_x-\omega-$spectrum \citep{goldschmidtTurbulentConvective81}:
\beq
u_c(y,z,\omega)
\equiv \frac{\omega}{k_{max}(y,z,\omega)}
\quad \text{with} \quad 
0 \equiv \frac{\partial}{\partial k_x} \, \tx{SPL}(k_x,y,z,\omega) \Big|_{k_x=k_{max}}.
\label{eq:uconv}
\eeq

\bcf[b]
\setlength\figureheight{0.55\textwidth}
\incg{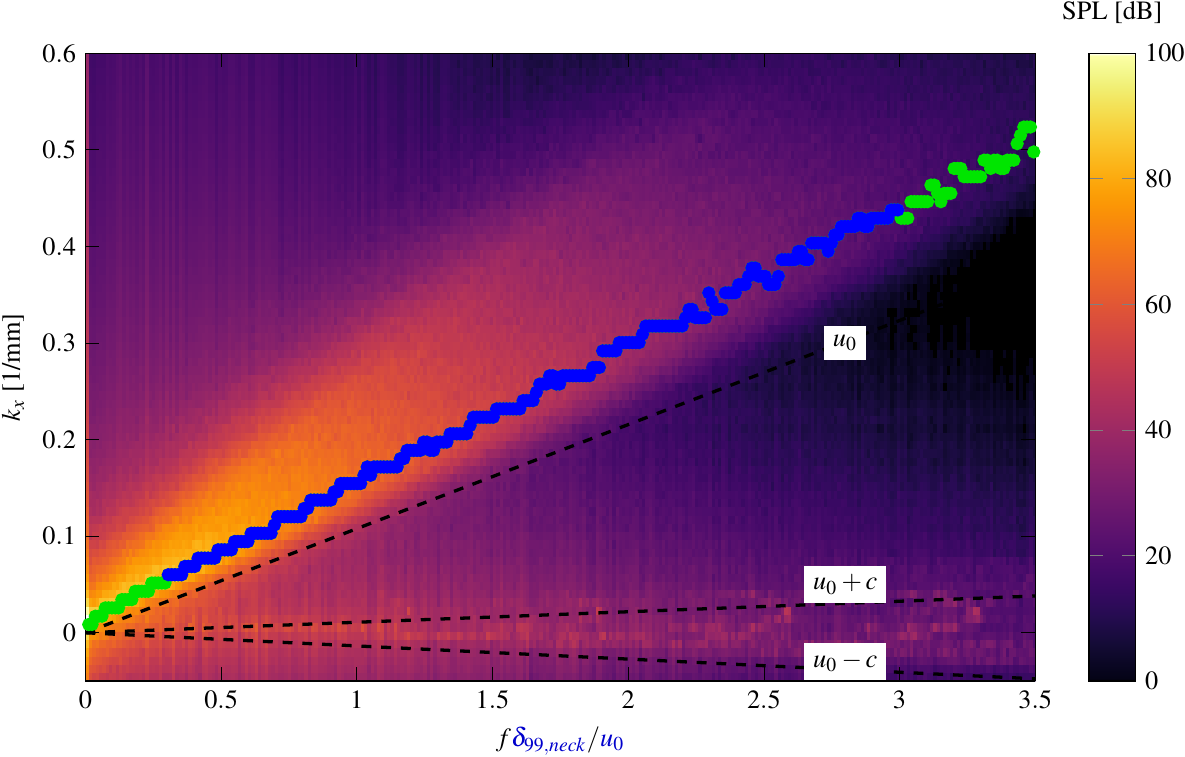}
\caption{
SPL $k_x-\omega-$spectrum %
of a turbulent flat-plate flow at $y=0$ ($z$-averaged, streamwise window of $13\gls{d99n}$). %
The characteristic phase velocities $\gls{u0}$ and $\gls{u0}\pm c$ are marked by dashed black lines.
The green and blue bullets denote the convective ridge.
Only the blue bullets are utilized later to define \gls{up} (\refe{updef}).
}
\label{fig:kf_spec_tbl}
\ecf

\bcf[t]
\setlength\figureheight{0.5\textwidth}
\incg{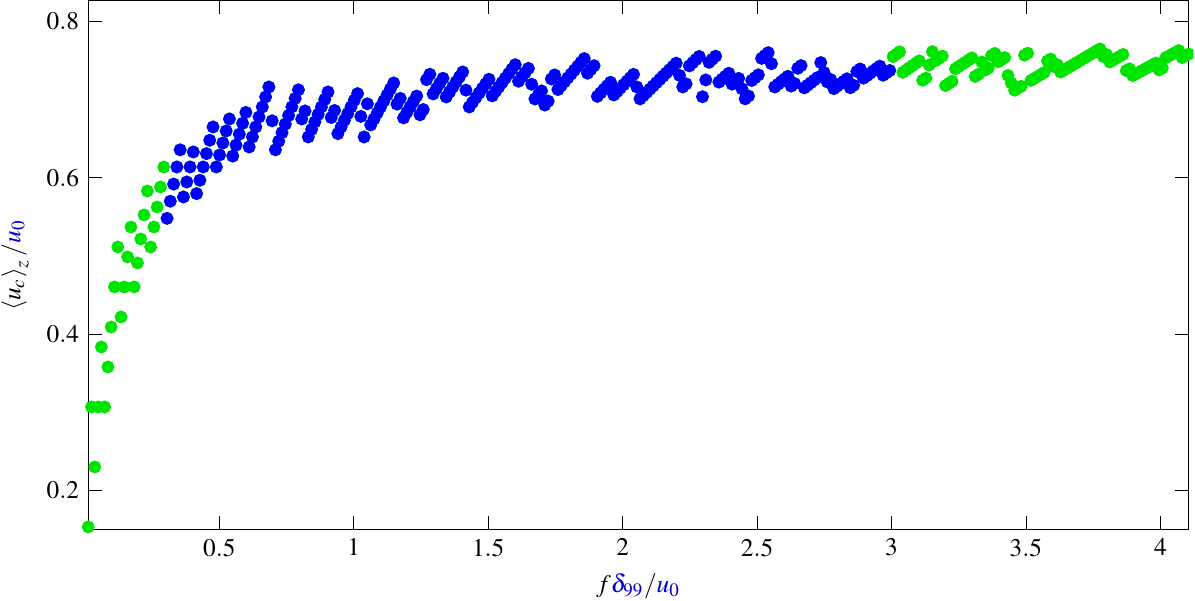}
\caption{
Convection velocity of a turbulent flat-plate flow calculated by \refe{uconv} ($z$-averaged, streamwise window of $13\gls{d99n}$ ($8L_{x-neck}$), $y=0$, same case as \reff{kf_spec_tbl}). Above the typical maximal curvature at $\omega\delta^*/\gls{u0} \approx 0.3$ the convection velocity is approximately constant.
Below, \gls{up} is defined as average from $\omega\delta^*/\gls{u0} =0.3$ to $\omega\delta^*/\gls{u0} =3$ (denoted by blue bullets).
}
\label{fig:uconv_over_f}
\ecf

Below, we will demonstrate how to apply \refe{uconv} on the $k_x-\omega-$spectrum provided by our DNS of the undisturbed TBL (s. previous \refs{source}).
Subsequently, we will calculate the general convection velocity in the more complex case of a TBL with a cavity, which is needed to derive a new definition for \gls{up}.

In \reff{kf_spec_tbl} the full DNS spectrum of the undisturbed TBL is shown. %
This more detailed $k_x-\omega-$representation separates acoustical and fluid-related (so-called convective) contributions.
All acoustic fluctuations are located around the frequency axis between the phase velocities $\gls{u0}\pm c$, while most fluid related pressure fluctuations are centered around the convective ridge with a phase velocity below the free stream velocity $\gls{u0}$.
Inside the convective ridge, most energy is stored, which is also reflected by the highest sound pressure levels.
The maxima of the convective ridge, which define the general convection velocity $u_c$ (\refe{uconv}), are marked by blue and green bullets.
In the interest of a clearer presentation, these maxima are replotted in \reff{uconv_over_f}.
At around $\omega\delta^*/\gls{u0} \approx f\gls{d99}/\gls{u0} \approx 0.3$ the convection velocity over the frequency exhibits a typical maximal curvature in accordance with 
\citet{viazzoSpectralFeatures01,gloerfeltTurbulentBoundarylayer13} and \citet{huNumericalStudy16}, where $\delta^*$  is the displacement thickness. Above this maximal curvature, the convection velocity remains approximately constant.
In the following subsection, we will use this characteristic $z$-averaged shape of $u_c$ to define \gls{up}.

To accurately determine the maxima
the calculation of a narrow band $k_x-\omega-$spectrum requires a resolution which fully resolves the location of the convective ridge in time and space
(like our DNS \citep{steinNumericalSimulation18} or the experiment by \citet{arguillatMeasuredWavenumber10}).
Measurements which are based on only a few microphones are incorrect, due to the known frequency and scale dependencies (i.e. streamwise probe separation) of the convection velocity \citep{kimPropagationVelocity93,gloerfeltTurbulentBoundarylayer13}.
Integral definitions of the convection velocity like in \citet{alamoEstimationTurbulent09} are not recommended in the present case of a distorted TBL since the convective ridge does not decay rapidly at high $|k_x|$ or $\omega$. This implies that the integral definition depends on the integration limits (of $k_x$ or $\omega$), which in turn depend on the data sampling rate.

In case a cavity is present below the general convection velocity $u_c$ can be equally calculated.
However, the acoustic range (defined between $\gls{u0}+c$ and $\gls{u0}-c$) should be excluded from the determination of the maximum $k_{max}$ (\refe{uconv}),
because localized SPL peaks of cavity modes can exceed the SPL of the convective ridge.
\\

$\bm{u_{+}}$ \textbf{Definition}\quad Based on the characteristic frequency dependence of the general convection velocity $\left< u_c \right>_{z}$,
the first step to spectrally define \gls{up} is to average $\left< u_c \right>_{Sr,z}$ between $\omega\delta^*/\gls{u0} =0.3$ and $\omega\delta^*/\gls{u0} =3$ (denoted by blue bullets).
Just the upper bound is arbitrary.
At least a shift of the upper bound has a marginal influence on the convection velocity
because $\left< u_c \right>_{z}$ is roughly constant around $Sr=3$ (s. \reff{uconv_over_f}) and decays only slowly at high frequencies far beyond $Sr=3$.
Beside this spectral determination of the convection velocity, $\left< u_c \right>_{Sr}$ will be also spatially confined to define \gls{up}.
Typically, larger TBL streaks occur further away from the no-slip wall, in faster moving fluid regions.
Consequently, the convection velocity increases with the streamwise length of the input signal $L_{x-DFT}$.
An asymptotic value of the frequency and $z$-averaged convection velocity $\left<u_c\right>_{Sr,z}/\gls{u0}\approx 0.7$
is reached as soon as the largest streaks of the TBL are captured by the streamwise signal window.
For the definition of $\gls{up}$,
only the structures which fit into the neck are of interest.
In the case of a TBL with a flush-mounted cavity,
the natural window length
to calculate the SPL $k_x-\omega-$spectrum
is the neck dimension.
Only inside the neck surface $S_{neck}$ the vortex sheet is modeled by Howe (\refe{vortex_disp}).
Hence, the streamwise DFT window is set equal to the streamwise neck length: $L_{x-DFT}=L_{x-neck}$.
In the following, the notation $\left<u_c\right>$ implies an average over $\omega\delta^*/\gls{u0}\in[0.3,3.0]$ and over $S_{neck}$.
\begin{figure}[b]
\setlength\figureheight{0.55\textwidth}
\incg{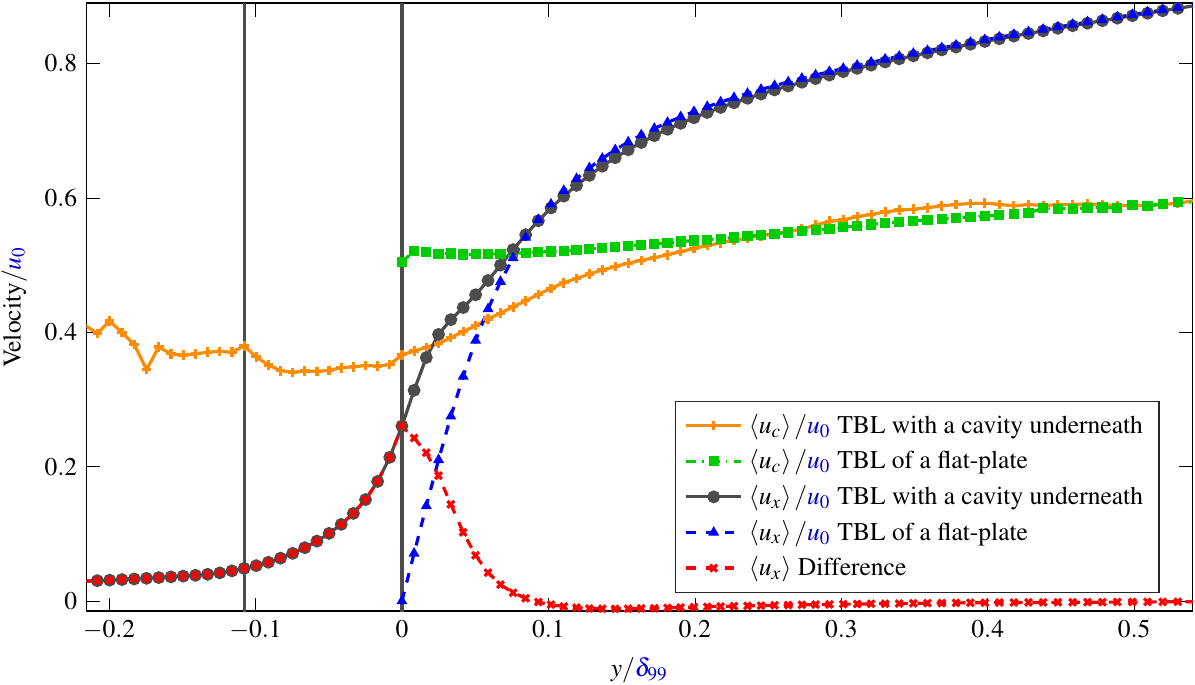}
\caption{Velocity profiles plotted over $y$ of a plane TBL and a TBL with a cavity underneath.
The notation $\left< \bullet \right>$ denotes an average over $\omega\delta^*/\gls{u0}\in[0.3,3.0]$ and over $x,z\in S_{neck}$.
The neck location is indicated by the vertical black lines between the cavity ceiling at $y=-1\si{mm}$ and the flat-plate surface at $y=0\si{mm}$.
}
\label{fig:u@neck}
\end{figure}
\bcf[h]
\setlength\figureheight{0.525\textwidth}
\incg{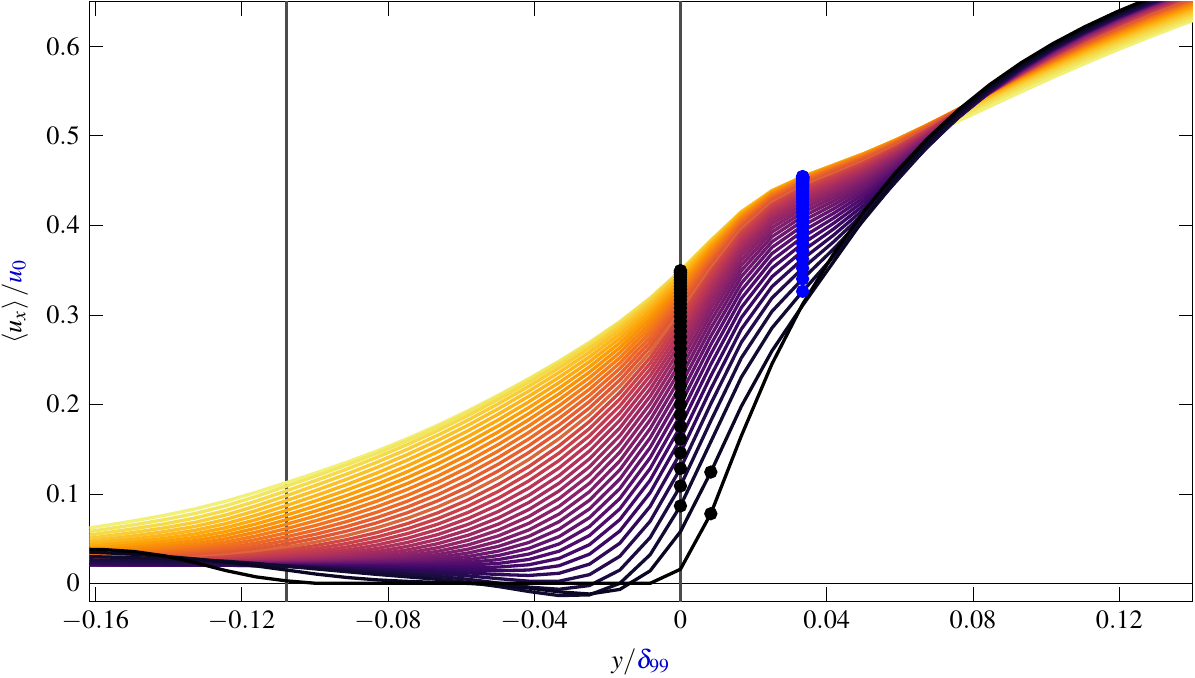}
\caption{
Indication of the constancy of the inflection point of the mean shear layer profile for different $x$ positions, starting with the leading edge of the neck at $197\si{mm}$ (black line) up to $209\si{mm}$ (yellow line).
All profiles are $z$-averaged.
Again the neck is aligned $y$-wise between $y=-1\si{mm}$ and $y=0\si{mm}$.
The first inflection point (black bullets) is located at the wall $y=0$ (vertical black lines).
The second inflection point (blue bullets) is located at $y^+\approx6$.
}
\label{fig:inflection}
\ecf

So far, the convection velocity $\left<u_c \right>$, which is averaged over $x$, $z$, and $\omega$, still depends on $y$.
In the following, this $y$ dependency of $\left<u_c \right>$ is examined,
before the characteristic height $y_+$ is selected, which ultimately defines $\gls{up} = \left<u_c \right> (y=y_+)$.
For this purpose, the DNS data of our reference case (s. \refs{case}) is further evaluated.

In \reff{u@neck} $\left<u_c \right>(y)$ is contrasted with the mean velocity $\left<u_x\right>(y)$.
For comparison, the same quantities of a plain TBL without a cavity are charted, too.
The mean velocity $\left<u_x\right>$ of the shear layer at the opening (cf. \reff{domain_opening_vsheet}) is greater than the plain TBL profile near the wall.
Only for $y>0.1\gls{d99n}$ both cases, with and without mounted cavity, coincide (consistent with the analytical Van-Driest profile \citep{steinNumericalSimulation18}).
The convection velocity $\left<u_c\right>$ increases with increasing wall distance, as noted before.
Remarkably, $\left<u_c\right>$ even exceeds $\left<u_x\right>$ near the wall in the viscous sublayer.
A possible explanation is that $u_c$ is a phase velocity (s. \refe{uconv}) and not a group velocity.
In the study of \cite{alamoEstimationTurbulent09}, the same behavior appears.
By comparing the case with an opening to the undisturbed TBL, we find an even lower $\left<u_c\right>$, despite a higher $\left<u_x\right>$.
We can assume that the newly formed eddies of the upstream edge have to be accelerated first so that the average (eddy) convection velocity drops.

In order to identify a characteristic height $y_+$,
we show in \reff{inflection} a whole series of boundary layer profiles ($z$ and time-averaged) at different streamwise neck positions, starting with the upstream edge.
The colormap varies from black to yellow with increasing $x$.
Beautifully, the first two inflection points of the profile have a constant height $y$ for all streamwise $x$ positions within the accuracy of the DNS mesh with $d x^+=1.5$. An upper "$+$" indicates \gls{dnu} normalization.
A third top inflection point scattering around $y/\gls{d99}=0.08$ is not marked to keep the \reff{inflection} clear.
The curvature of the two outer inflection points (first and third) meets the Fjørtoft's criterion necessary for a shear layer instability \citep{schmidStabilityTransition12} in contrast to the second inner inflection point.
To conclude the determination of $\gls{up}$, we select the second central inflection point (located between the two others), as the characteristic location of the neck shear layer $y_+=y_\tx{central inflection pt.}$:
\beq
\gls{up} \equiv
\left<u_c \right>_{Sr,\,S_{neck}} (y_\tx{central inflection pt.})
\;,
\label{eq:updef}
\eeq
being $u_c$ defined by \refe{uconv} and averaged over $S_{neck}$ and $\omega\delta^*/\gls{u0}\in[0.3,3.0]$.
In the present case the point of inflection is $y_+^+=6\pm1.5$ ($y_+=0.31\pm0.08\si{mm}$),
which results according to \reff{u@neck} in $\gls{up}/\gls{u0}=0.39\pm0.03$.
\gls{up} directly scales with \gls{u0} here.
This \gls{u0} dependency is in agreement with experiments carried out at various velocities $\gls{u0} \in[5\si{m/s},47\si{m/s}]$
(see the following \reff{howe}, \reff{model_over_u}{exp_over_u} in \refs{results}).
Thus, we set up a robust $\gls{up}$
definition.
\\

$\bm{\beta}$ \textbf{Definition}\quad
As the most obvious approach, corresponding to the common original interpretation of $u_\pm$, we set
$\gls{beta}=u_-/\gls{up}$ to be proportional to the mean velocities $\left<u_x\right>$ below and above the opening.
Though, we will clarify in the following, that this choice is incapable to describe a realistic opening with a vortex sheet.
Subsequently, an alternative definition is suggested.
In the present case of one-sided flow, the obvious (but discarded) approach results in the hypothetical proportional relation
$
\gls{beta}
=
\left<u_x\right>(y=\delta_-^*)/\left<u_x\right>(y=\delta_+^*)
\propto
1/\gls{u0}
\;,
$
where $\delta_\pm^*$ represent a measure of the shear layer thickness above and below the opening.
Since $\delta_\pm^*$ typically grows linearly in streamwise $x$ direction \cite[Sect. 5.4.2]{popeTurbulentFlows00}, $\gls{beta}$ is roughly $x$ independent.
$\delta_+^*$ is approximated as the location, where the difference between the shear layer profile and the plane TBL profile falls until $0.1\gls{u0}$.
Furthermore, we set $\delta_-^*=\delta_+^*$.
In case of the DNS data with $\gls{u0}=\SI{38.5}{m/s}$ we get $\gls{beta}=8/\gls{u0}=0.21$. %
However, this $\gls{u0}$ dependency is not reflected by experiments.
As an alternative, the best fit of $Z_{flow}$ (see the following \refs{ZflowValid}) and the complete Helmholtz resonator model (\refs{range}) with Golliard's experiments is achieved with a constant
\beq
\gls{beta}\equiv 0.21
\label{eq:umdef}
\eeq
relation, which is valid for different $\gls{u0} \in[5\si{m/s},47\si{m/s}]$ (s. \reff{model_over_u}{exp_over_u}).
This definition of $\gls{beta}$ remains to be further investigated (more comments follow in \refs{ZflowValid} and \refs{reinterpretation}).

\section{Results: Interpretation, Validation, and Validity Range of the Model}
\label{sec:results}

In this section, the Helmholtz resonator model with its new parameters is validated by comparison with an experiment and our DNS:
First, the comparison is started with the most crucial model element $Z_{flow}(\gls{up},\gls{beta})$, separately (\refs{ZflowValid}).
Second, a new physical meaning of the governing model parameters \gls{up} and \gls{beta} is revealed (\refs{reinterpretation}).
Third, the complete Helmholtz resonator model is compared with our DNS results (\refs{valid}).
Fourth, the validity range of the new model is demonstrated (\refs{range}).

\subsection{Validation of the Flow-Acoustic Interaction Impedance $Z_{flow}(\gls{up},\gls{beta})$}
\label{sec:ZflowValid}

\bcf[t]
\setlength\figureheight{0.5\textwidth}
\incg{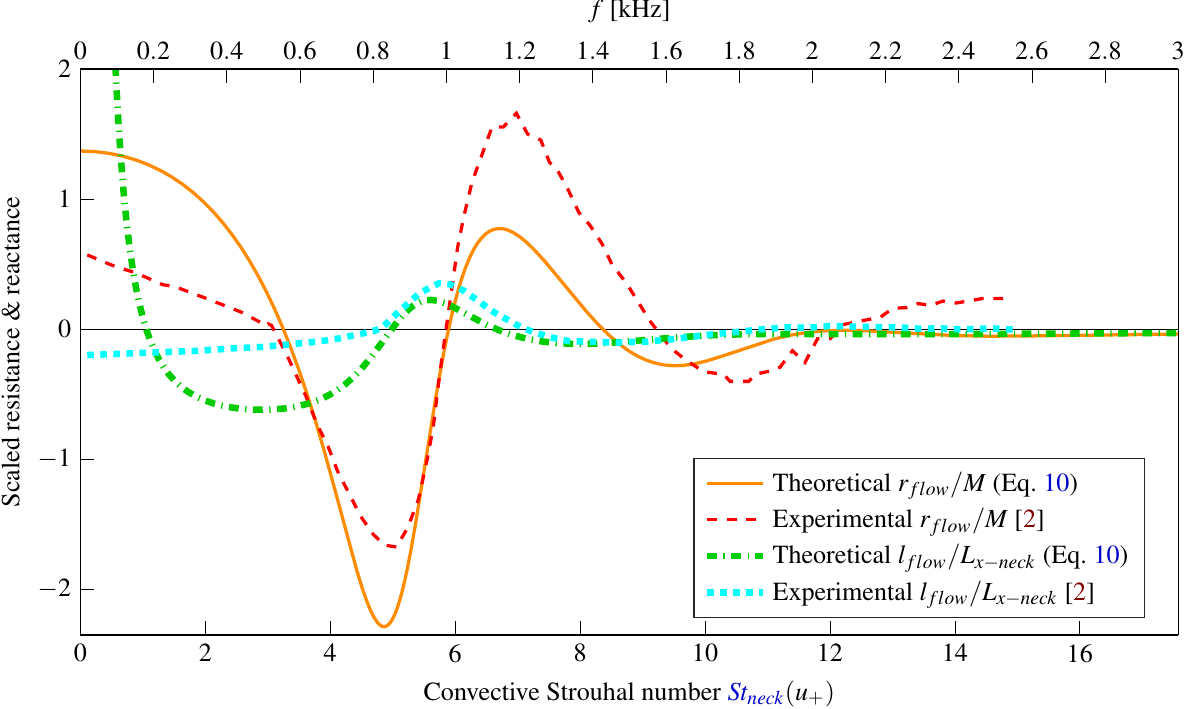}
\caption{
Validation of Howe's impedance model \refe{Zflow} vs. experiment for $\gls{up}/\gls{u0}=0.39$ and $\gls{beta}=0.21$ (s. \refs{uconv}).
The resistance $r_{flow}$ and the reactance $l_{flow}$ are rescaled with the Mach number $M$ and the streamwise neck length $L_{x-neck}$, respectively.
}
\label{fig:howe}
\ecf

\reff{howe} shows the impedance $Z_{flow}/Y_{neck}=r_{flow}+i k l_{flow}$ modeled by \refe{Zflow} %
and measurements %
by %
\cite{golliardNoiseHelmholtzresonator02}.
Thereby the experimental curves are averaged over different conditions such as $M\in[0.11,0.12]$ and ratios of $\gls{d99n}/L_{x-neck}\in[0.7,5]$.
Even though \citeauthor{howeAcousticsFluidStructure98}'s theoretical model assumes an infinite thin vortex sheet, the experimental results are in accordance with the model predictions:
especially the frequency scaling i.e. the extrema positions and the zero-crossings match.
At the zero-flow limit (\gls{Sneck} at infinity) the impedance becomes zero, as expected: $\lim_{\gls{Sneck} \to \infty} Z_{flow} = 0$.
The sign of $r_{flow}$ determines the direction of the energy transfer between the Kelvin-Helmholtz waves of the neck and the passing sound waves.
The acoustic damping at low Strouhal numbers $\gls{Sneck}<3.3$, as well as the acoustic excitation between $\gls{Sneck}\in[3.3,5.9]$, is typical for all kind of neck geometries like circles, rectangles or triangles \citep{Grace1998a}.
The coupling of Kelvin-Helmholtz waves and acoustic waves can be viewed similarly to \citeauthor{Rossiter1964}'s feedback mechanism.
The crucial difference is that Rossiter's self-generated sound waves propagate upstream in the vortex alley,
while in the present model (\refs{howe}) external plane waves transversally pass through the vortex sheet.

The impact of $\gls{up}$ and $\gls{beta}$ on $Z_{flow}$ is as follows:
$\gls{up}$ determines the main frequency scaling, which relates the Strouhal number $\gls{Sneck}$ and the frequency $f$ (s. \refe{KHwavenumberSrBeta}).
If $\gls{up}$ is modified, the entire \reff{howe} remains unchanged with the sole exception of the upper frequency axis.
Such $Z_{flow}$ plotted over $\gls{Sneck}$ (as in \reff{howe}) allows a universal representation.
\gls{beta} tunes the amplitude of the complex-valued impedance $Z_{flow}$ depending on the frequency. %
If $\gls{beta}$ is increased from zero to one,
constantly new positive and negative bulges of $r_{flow}$ and $l_{flow}$ arise at higher and higher frequencies \citep[cf.][Fig.~5.3.8]{howeAcousticsFluidStructure98}.
In doing so, the new bulges have a slowly, shrinking amplitude with higher frequency,
while the amplitude and position of the low-frequency bulges remain almost the same.
Here, our definition of $\gls{beta}=0.21$ by \refe{umdef}
overestimates low-frequency and underestimates high-frequency amplitudes, as shown in \reff{howe}.
This discrepancy is an indication that the single parameter $\gls{beta}$ of Howe's theory
is not sufficient to accurately describe low and high-frequency amplitudes simultaneously.

\subsection{A New View on the Model Parameters \gls{up} and \gls{beta}}
\label{sec:reinterpretation}

In conclusion of \refs{Zflow} and \refs{ZflowValid}, a reinterpretation of \gls{up} and \gls{beta} suggests itself.
We propose to consider \gls{up} as the central vortex sheet velocity,
i.e. the convection velocity at the central inflection point of the neck shear layer.
Hence \gls{up} is a central quantity of the vortex sheet, rather than a velocity located above the vortex sheet as suggested by the lower ``$+$'' index.
In the following, the plus sign is kept to remind of its origin within Howe's theory.

Instead of understanding \gls{beta} only as a velocity ratio $u_-/\gls{up}$ (\refe{KHwavenumberSrBeta}),
\gls{beta} should be more universally regarded as a measure of the shear layer growth within the opening.
Other than the mean velocity difference below and above the vortex sheet $\left<u_x\right>_\pm$ (original interpretation),
the growth of the shear layer thickness is likely to be affected by the opening thickness $L_{y-neck}$ %
and the relative, incoming boundary layer thickness $\gls{d99}/L_{x-neck}$.
The motivation to regard \gls{beta} as a general measure of the shear layer growth is as follows:
In the limit $\gls{beta}=1$, $Z_{flow}$ (s. \reff{howe}) has contributions (bulges) over the whole frequency range
(and a purely imaginary-valued Kelvin-Helmholtz wave exponent of \refe{vortex_disp}).
If high frequencies have a larger influence, small structures are more important, which indicates a thin vortex sheet and a minimal shear layer growth.
By contrast, in the limit $\gls{beta}=0$, $Z_{flow}$ has only low-frequency components, large structures are dominant, which requires a rapid shear layer growth within the streamwise opening length.
\citet[Sect. 5.3.6]{howeAcousticsFluidStructure98} himself argues that the conductivity of an opening in two-sided flow, in practice, is actually similar to the case of one-sided flow, because of the finite $L_{y-neck}$ thickness (not infinitely thin as theoretically assumed).
In other words, the recirculation area (caused by thicker plate) at the upstream side of the opening enhances the shear layer growth and therefore can be represented by an effectively smaller $\gls{beta}$.
To sum up, \gls{beta} is (at least) a function of $\left<u_x\right>_\pm$, $L_{y-neck}$, and $\gls{d99}/L_{x-neck}$.
It remains to prove and explicitly quantify these %
predictions about the qualitative behavior of \gls{beta}. %

\subsection{Validation of the Helmholtz Resonator Model}
\label{sec:valid}

\bcf[b]
\setlength\figureheight{0.55\textwidth}
\incg{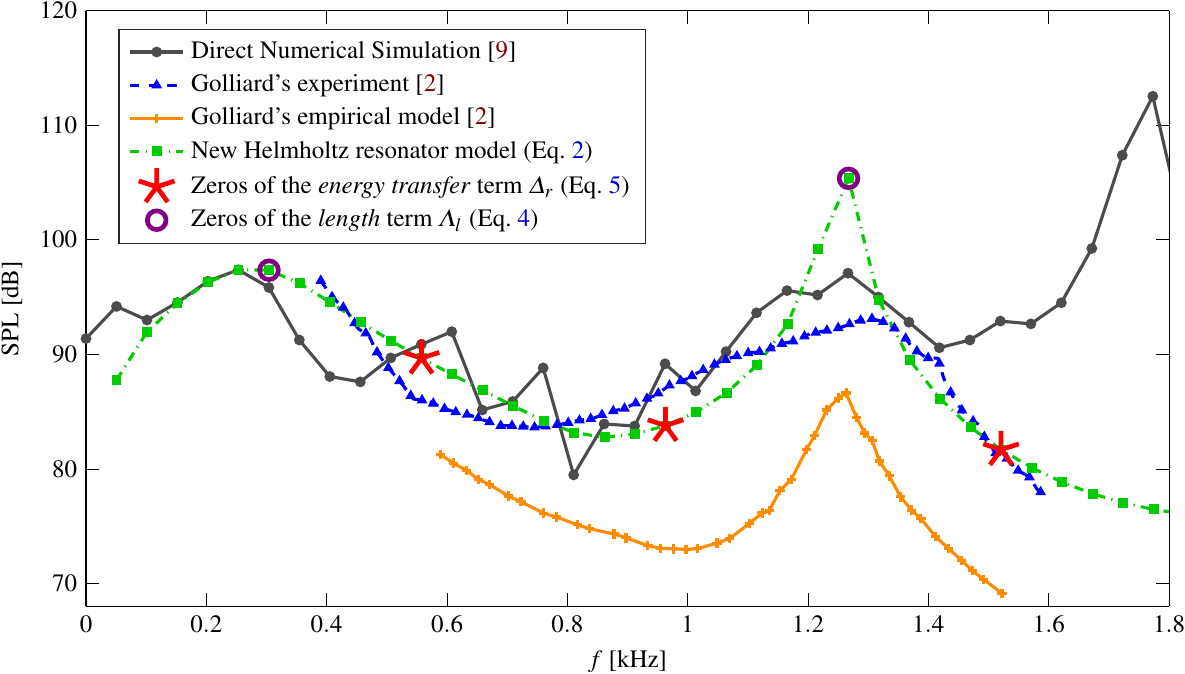}
\caption{
Narrow-band SPL spectrum ($f_{bin}=\SI{50}{\Hz}$) centered at the bottom of the Helmholtz cavity (setup s. \reff{setup}).
For comparison, the measured SPL and the empirical model of \cite{golliardNoiseHelmholtzresonator02} are shown.
$x$- and $z$-cavity modes, which are not part of the model, begin at \SI{1700}{\Hz}.
This explains the deviation starting near this frequency.
}
\label{fig:modelVSdns}
\ecf

The application of the complete Helmholtz resonator model (\refe{GHGmodel} with \refe{HRtransfer}{goody}) is presented in \reff{modelVSdns}.
Here the SPL narrowband spectrum at the bottom of the cavity is shown.
At the bottom of a cavity unsteady, acoustical pressure fluctuations dominate (cf. \reff{domain_opening_vsheet}).
This has the advantage that without a Helmholtz decomposition the power spectral density $\Phi_{bot}$ of the pressure is directly related to the SPL.
As a comparison, the $xz$ averaged SPL of the DNS and the experimental SPL of \citet[cavity C1, TBL B, $L_{x-neck}=14\si{mm}$, $\gls{d99}=10.9\si{mm}$]{golliardNoiseHelmholtzresonator02} are plotted.
Also, the model predictions of an alternative empirical model are taken from \citet[Sect. 5.3.1]{golliardNoiseHelmholtzresonator02}.
The predicted extrema frequencies of the Helmholtz resonator model match with the experimental and the DNS results.
Zeros of the \emph{length} term $\Lambda_l$ (\refe{dtot}) are denoted by violet circles:
The first SPL peak of \reff{modelVSdns} at \SI{300}{\Hz} is the Helmholtz base frequency.
The second peak at \SI{1200}{\Hz} is the first vertical $y$-cavity mode.
Below \SI{1500}{\Hz} the SPL of the new model, the experiment, and the DNS deviates by $\pm 7 $dB. %
$x$- or $z$-cavity modes begin at \SI{1700}{\Hz} ($c/2 L_{x-cavity}$, where $c$ is the speed of sound).
Hence, the negative deviation of the model at \SI{1700}{\Hz} is expected.
By virtue of simplicity, these higher transverse $x$- or $z$-modes are not included in the model because they are beyond the typical range of operation.
Zeros of the \emph{energy transfer} term $\Delta_{r}$ (\refe{rtot}) are denoted by red stars.
Here the interchange between acoustical and fluid energy is balanced over the time average.
Since the ratio $S_{ratio}=S_{neck}/S_{cavity}=0.14$ of \refe{HRtransfer} is small, they hardly contribute to the overall SPL at the cavity bottom.
Only in the DNS results, small local increases are visible at zero resistance (red stars in \reff{modelVSdns}).
However, this might be up to the noise due to the limited time series available of $20 \si{ms}$ (restriction of DNS resources).

Golliard's empirical model is only evaluated within its frequencies validity range.
Thereby, the frequency prediction of the first vertical $y$-cavity mode is in agreement with the other curves, however, a constant negative offset of about $-10$dB appears.

\subsection{Validity Range of the Helmholtz Resonator Model}
\label{sec:range}

\bcf[p]
\setlength\figureheight{0.52\textwidth}
\incg{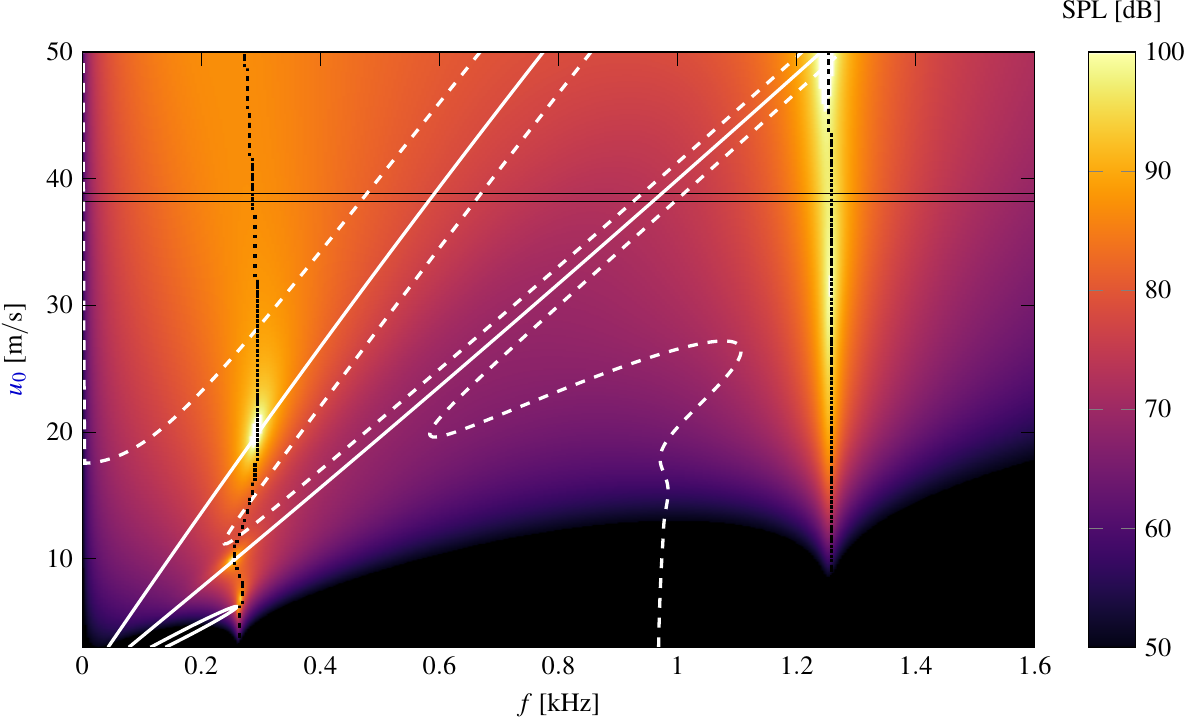}
\caption{
SPL spectrum ($f_{bin}=\SI{4.3}{\Hz}$) predicted by the new Helmholtz resonator model \refe{GHGmodel} for the cavity setup of \reff{setup}.
The dotted black vertical lines mark the resonance conditions i.e. the \emph{length} term zeros $\Lambda_l=0$ (\refe{dtot}).
The white lines mark $r_{flow}+r_{rad}=0$ and the dashed white lines mark $|r_{flow}+r_{rad}|=0.07$.
Consequently, both white lines are related to the \emph{energy transfer} term \refe{rtot}.
The black horizontal double line corresponds to the DNS case with $\gls{u0}=38.5\si{m/s}$ (s. \reff{modelVSdns}).
}
\label{fig:model_over_u}
%
\setlength\figureheight{0.52\textwidth}
\incg{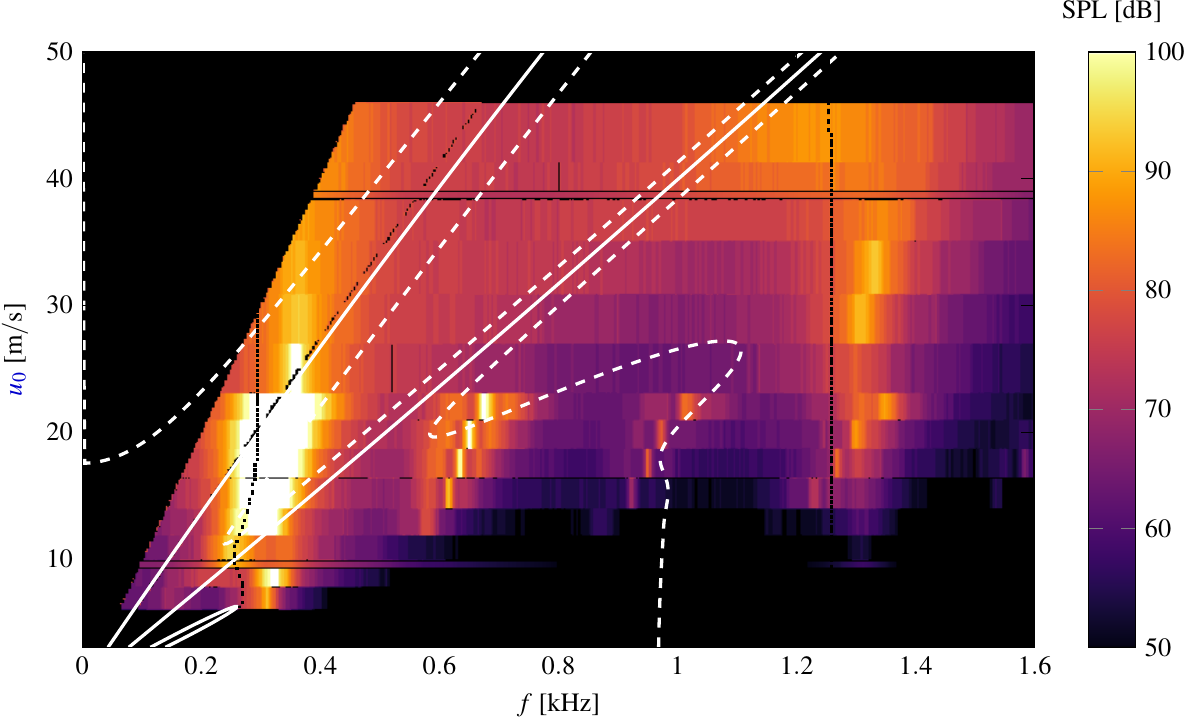}
\caption{
Measured SPL spectrum extracted from \cite{golliardNoiseHelmholtzresonator02} for the same conditions as in \reff{model_over_u} (effective $f_{bin}=\SI{4.3}{\Hz}$, original $f_{bin,exp}=\SI{1}{\Hz}$).
The dotted black and the white lines are a copy from \reff{model_over_u}.
Lines which are only present in \reff{exp_over_u} are artifacts caused by the data extraction.}
\label{fig:exp_over_u}
\ecf

During the derivation of the model, the focus is on universal definitions of the governing parameters.
However, the reference data is provided by one single DNS run with $\gls{u0}=38.5\si{m/s}$ (\refs{case}).
To demonstrate its universal validity, the model is evaluated for the whole velocity range from $5\si{m/s}$ to $47\si{m/s}$ ($M\in[0.01,0.14]$) measured by \citet{golliardNoiseHelmholtzresonator02}.
In \reff{model_over_u}{exp_over_u} the SPL model spectra can be compared with the experimental measurements, respectively.
To a great extent, both figures are in agreement at all velocities.

The zeros of the \emph{effective length} $\Lambda_l$ (\refe{dtot}, dotted black mainly vertical lines) are predominantly geometry dependent so that the position of the Helmholtz base frequency (at \SI{304}{\Hz}) and the
first vertical $y$-cavity mode (at \SI{1154}{\Hz}) are almost velocity independent.
In contrast, the zeros of the \emph{energy transfer} $\Delta_r$ (\refe{rtot}, dashed white diagonal lines), which correspond to a balanced energy interchange between the acoustical and the Kelvin-Helmholtz waves, scale as expected with the free stream velocity $\gls{u0}$.
The overlap of both resonance conditions leads to a strong nonlinear interaction of the vortex-sheet with the cavity modes.
A qualitative similar behavior of these resonance conditions is also observed for example by Yang's \citep{Yang2009} experimental study of deep, open cavities (without a neck).
The new Helmholtz resonator model successfully predicts where the occurring physical phenomena are coupled:
the Helmholtz base frequency strongly couples with the Kelvin-Helmholtz waves of the neck between $\SI{5}{m/s}$ and $\SI{23}{m/s}$,
resulting in a SPL beyond 120\si{dB} (s. \reff{model_over_u}).
It is known that such high sound pressure levels cause a scattering to higher harmonics \citep{fornerScatteringHigher17}.
Experimentally these higher harmonics are visible in \reff{exp_over_u} at 330, 660, 990 and 1320 \si{Hz}.
The model predicts the initial resonance condition at $\approx \SI{300}{Hz}$ and $\gls{u0} \in [\SI{5}{m/s},\SI{23}{m/s}]$, but no scattered higher harmonics
(nonlinear effect).
The right prediction of the resonance conditions by $\Lambda_{\delta}(\gls{up},\gls{beta})$ and $\Delta_r(\gls{up},\gls{beta})$ for the frequencies $f\in[0,\SI{1500}{Hz}]$ and for the Mach numbers $M\in[0.01,0.14]$ verifies the new \gls{up} (\refe{updef}) and \gls{beta} (\refe{umdef}) definitions, within these ranges.

\section{Model Summary}
\label{sec:model_sum}

The Helmholtz resonator model as given by \refe{HRtransfer}{goody} with its generalized parameters $\gls{up}$ and $\gls{beta}$ is ready for use.
An industrial user can directly apply it, by evaluating
\beq
\Phi_{bot}=T(Z_{neck},S_{neck},S_{cavity},k\,L_{y-cavity})\, \Phi_{TBL}(\omega, \gls{d99n}, \gls{u0}, \gls{utau}, \rho_w, \nu).\notag
\eeq
Thereby, the neck impedance is given in detail by
\beq
Z_{neck} =
Z_{jump}(S_{neck},S_{cavity}) +
Z_{flow}(L_{x-neck},\gls{Sneck},\gls{beta}) +
Z_{rad}(L_{x-neck},L_{z-neck}),\notag
\eeq
where the individual impedance elements originate from \refe{howeUzeroD}{Zflow}.
The new $\gls{up}$ is determined as follows:
First, the $k_x-\omega-$spectrum is calculated exactly above the neck surface at a constant height of the central inflection point of the shear layer.
Second, the convection velocity $u_c$ is derived as the phase velocity $\omega/k_x$ of the pressure $k_x-\omega-$spectrum at the convective ridge (maximum).
Third, $\gls{up}$ is the mean of $u_c(f)$ for $f\gls{d99n}/\gls{u0} \in[0.3,3.0]$ and for all spanwise neck positions (\refe{updef}).
In the present case of one-sided flow, $\gls{beta}$ is defined as constant $0.21$ (\refe{umdef}).

The model is validated
for low frequencies below the cutoff of higher transverse cavity modes ($f\in[0,\SI{1500}{Hz}]$) and
for low Mach numbers from $0.01$ to $0.14$ (\refs{range}).
These ranges comply with typical operating conditions of duct systems.
The model is based on a modular design (\refe{rawmod}), which guarantees a flexible use and expandability.

\section{Conclusion}
\label{sec:end}

Altogether, by combining and carrying on works by Goody, Howe, and Golliard we established a new model of a Helmholtz resonator in grazing flow.
The derivation of the model is based on DNS data.
Goody's model is selected as the most realistic acoustic source term of a TBL and adapted to our model (\refs{source}).

The physical key point of this paper is the correct description of the turbulence-acoustic interaction.
The understanding of the related energy interchange is vital to predicting acoustical damping or excitation.
To this end, Howe's extensive theory of this interaction is put on a new physical footing and incorporated in the new model.

The main contribution is the novel definition and reinterpretation of the parameters $\gls{up}$ and $\gls{beta}$,
which are governing Howe's theory.
So far, these parameters are empirically fitted depending on the particular application.
Now $\gls{up}$ and $\gls{beta}$ are uniquely defined and directly related to physically significant quantities:
$\gls{up}$ is the central, temporal and spatial specified convection velocity of the neck shear layer.
$1/\gls{beta}$ is a measure of the shear layer growth (\refs{reinterpretation}).

Because the turbulence-acoustic interaction is based on Howe's simplified impedance model (s. \refs{howe}),
some restriction apply:
Effects caused by a modified wall thicknesses $L_{y-neck}$
or induced by an altered boundary layer thickness \gls{d99}
require to set up an extended database and to re-evaluate
$\gls{up}$, $\gls{beta}$
by using their new universal definitions derived here
(continuation option of this work).

By applying the new $\gls{up}$, $\gls{beta}$ definitions to the DNS data and by comparison with experiments by other researchers it is shown
that changes in the frequency or the free stream velocity (\refs{range})
have an appropriate impact on the final model predictions.
Different free stream velocities \gls{u0}, frequencies $f$ and most geometry modifications\footnote{\refe{vortex_disp} can be solved for any $S_{neck}(x,z)$ shape.}
can be studied within the ranges $f\in[0,\SI{1500}{Hz}]$, $M\in[0.01,0.14]$.
Hence, the change of outer parameters is now traceably linked to governing parameters, which in turn directly update the model predictions.
For instance, a larger free stream velocity \gls{u0} %
increases the convection velocity $\gls{up}$.
The updated Strouhal number $\gls{Sneck}(\gls{up})$ then newly determines at which frequencies acoustic energy is damped or excited (sign of $\Delta_r$, \refe{rtot})
and restates the resonance condition of the instability waves of the opening and the cavity modes depending on the frequency and the velocity \gls{u0} (cf. \reff{model_over_u}).
A priori, rather than by expensive tests, the sound spectrum can be directly tuned for frequencies of interest.

\begin{acknowledgements}
We gratefully acknowledge the financial support of the Deutsche Forschungsgemeinschaft (SE824/29-1) and
the provision of computational resources by the High-Performance Computing Center Stuttgart (ACID11700).
\end{acknowledgements}

\bibliography{bibtex}   %

\end{document}